\documentclass[floatfix,
 reprint,
superscriptaddress,
preprintnumbers,
nofootinbib,
 amsmath,amssymb,
 aps
]{revtex4-1}

\usepackage[colorlinks]{hyperref}
\usepackage{graphicx}
\usepackage{dcolumn}
\usepackage{bm}
\usepackage{color}
\usepackage{placeins}
\usepackage{slashed}
\usepackage{stackrel}
\usepackage{multirow}
\usepackage{tabularx}
\usepackage{mathtools}
\usepackage{floatrow}
\usepackage{blindtext}
\usepackage{cancel}
\usepackage{enumitem}
\usepackage{siunitx}
\usepackage{dsfont}
\usepackage[capitalise, english]{cleveref}

\AtBeginDocument{\usepackage{booktabs}}

\usepackage{array}
\newcolumntype{L}[1]{>{\raggedright\let\newline\\\arraybackslash\hspace{0pt}}m{#1}}
\newcolumntype{C}[1]{>{\centering\let\newline\\\arraybackslash\hspace{0pt}}m{#1}}
\newcolumntype{R}[1]{>{\raggedleft\let\newline\\\arraybackslash\hspace{0pt}}m{#1}}

\definecolor{red}{rgb}{0.9, 0,0}
\definecolor{cerulean}{rgb}{0., 0.62,0.9}
\definecolor{navy}{rgb}{0.05, 0.05,0.8}

\DeclareSIUnit\pc{\text{pc}}

\usepackage[usenames,dvipsnames,svgnames,table]{xcolor}

\def\be{\begin{equation}}
\def\ee{\end{equation}}
\def\bea{\begin{eqnarray}}
\def\eea{\end{eqnarray}}

\newcommand{\mdm}{M_{\chi}}
\newcommand{\vrel}{v}

\newcommand{\shard}{\sigma^h}
\newcommand{\V}{\tilde{V}}

\newcommand{\chis}{\phi_{j,i}}
\newcommand{\chibs}{\phi_{nlm}}
\newcommand{\vlo}{V^{\text{LO}}}

\newcommand{\xcut}{x_{\rm cut}}
\newcommand{\vcut}{\mathcal{V}_{\rm cut}}

\newcommand{\schr}{Schr\"{o}edinger~}

\def\beq{\begin{equation}}
\def\eeq{\end{equation}}

\def\be{\begin{eqnarray}}
\def\ee{\end{eqnarray}}

\begin{document}
\widetext

\title{The Sommerfeld enhancement at NLO and the dark matter unitarity bound}
\author{Salvatore Bottaro}
\affiliation{Raymond and Beverly Sackler School of Physics and Astronomy, Tel-Aviv 69978, Israel}
\author{Diego Redigolo}
\affiliation{INFN, Sezione di Firenze Via G. Sansone 1, 50019 Sesto Fiorentino, Italy}

\begin{abstract}
We reexamine the consequences of perturbative unitarity on dark matter freeze-out when both Sommerfeld enhancement and bound state formation affect dark matter annihilations. At leading order (LO) the annihilation cross-section is infrared dominated and the connection between the unitarity bound and the upper bound on the dark matter mass depends only on how the different partial waves are populated. We compute how this picture is modified at next-to-leading order (NLO) with the goal of assigning a reliable theory uncertainty to the freeze-out predictions. We explicitly compute NLO corrections in a simple model with abelian gauge interactions and provide an estimate of the theoretical uncertainty for the thermal masses of heavy electroweak $n$-plets. Along the way, we clarify the regularization and matching procedure necessary to deal with singular potentials in quantum mechanics with a calculable, relativistic UV completion.    
\end{abstract}

\maketitle
\noindent


\section{Introduction}

Among the plethora of dark matter (DM) production mechanisms, a minimal and predictive setup is DM thermal freeze-out where the DM is in thermal contact with the Standard Model (SM) bath in the early Universe, and its abundance today is set by 2-2 annihilations into SM or dark sector states. 

This simple framework makes it possible to derive an upper bound on the DM mass from the perturbative unitarity of the annihilation cross-section as was first done in Ref.~\cite{Griest:1989wd}. The unitarity of the S-matrix bounds from above every single partial wave contributing to the annihilation cross-section. At a given order in the perturbative expansion, this bound can be recast into a maximal value of the gauge coupling. The latter can then be translated into an upper bound on the DM mass \cite{Baldes:2017gzw} through the requirement that the total annihilation cross-section should deplete the DM abundance to match the measured relic density today. 

When long-range interactions are at work, non-relativistic (NR) quantum mechanical effects significantly alter the DM annihilation cross-section inducing an overall enhancement which is dubbed Sommerfeld enhancement (SE) in the literature~\cite{Hisano:2006nn,Cirelli:2007xd,Iengo:2009ni,Iengo:2009xf,Hryczuk:2011vi,Cassel:2009wt} typically accompanied by the bound state formation (BSF) during the annihilation process~\cite{Asadi:2016ybp,Petraki:2015hla,Harz:2018csl,Mitridate:2017izz,vonHarling:2014kha}. Since these effects dominate the annihilation cross-section it is crucial to understand their behavior once we approach the perturbative unitarity bound (PUB). This question is the main focus of this paper.

Approaching the perturbative unitarity bound, NLO corrections should become important and their relative size compared to the leading order contributions gives a reliable estimate of the expected theory uncertainty. NLO corrections to the DM annihilation cross-section have been studied extensively in the past in various models \cite{Baro:2009na,Chatterjee:2012hkk,Harz:2012fz,Hellmann:2013jxa,Ovanesyan:2016vkk,Schmiemann:2019czm,Biondini:2023zcz} (see 
\cite{Binder:2020efn,Binder:2021otw} instead for different BSF channels), in this study we focus on the behavior of NLO corrections to the  non-relativistic potential.  Making use of the general results from Ref.~\cite{Beneke:2013jia,Beneke:2019qaa,Beneke:2020vff}, we systematically include corrections generated both by the infrared (IR) and ultraviolet (UV) dynamics. 

Our analysis allows us to reliably assign a theoretical uncertainty to the freeze-out predictions. Even though our results are general, we use a simple dark QED model to illustrate the impact of NLO corrections. We will comment on how this analysis allows us to assign a more reliable theory error to the electroweak WIMPs thermal masses derived in Ref.~\cite{Bottaro:2022one,Bottaro:2021snn}.

Our paper is structured as follows. In Sec.~\ref{sec:DQED} we summarize the relevant ingredients for the leading order freeze-out computation and discuss the perturbative unitarity bound at leading order. In Sec.~\ref{sec:NLO} we develop the tools to account for both IR (Sec.~\ref{sub:IR}) and UV NLO corrections (Sec.~\ref{sec:UV}). In Sec.~\ref{sec:darkQED} we then illustrate the importance of our corrections in a simple dark QED model. In Sec.~\ref{sec:conclusions} we conclude. In Appendix~\ref{app:fourier} we illustrate the general strcture of UV NLO potentials while in Appendix~\ref{sec:coulomb} we illustrate our regularization and matching procedure in the simple case of the Coulomb potential. In Appendix~\ref{sec:BSFNLO} we collect useful formulas about bound state formation.

\section{The Unitarity bound at leading order}
\label{sec:DQED}

 We first discuss DM annihilation at leading order. We illustrate SE in Sec.~\ref{sec:hardLOandSE} and BSF in Sec.~\ref{subsec:BSFLO}. The DM perturbative unitarity bound at leading order and its consequences are considered in Sec.~\ref{subsec:uniLO}.  We decompose the annihilation channels in eigenvalues of the total angular momentum $\vec{J}=\vec{L}+\vec{S}$, where $\vec{L}$ is the angular momentum and $\vec{S}$ is the internal spin.\footnote{The Casimir operators are defined in the standard way: $\vec{J}^2=j(j+1)\mathds{1}$, $\vec{L}^2=l(l+1)\mathds{1}$, $\vec{S}^2=s(s+1)\mathds{1}$.} 

Since the freeze-out happens at non-relativistic velocities, the annihilation channels with $\vec{L}\neq0$ are velocity suppressed. We can then focus on s-wave annihilation and set $\vec{L}=0$ and $\vec{J}=\vec{S}$. We consider the DM to be a scalar or a fermion, where in the latter case both $j=1$ and $j=0$ channels contribute to the s-wave annihilation. For simplicity, we take the DM to be in thermal contact with the SM in the early Universe. As a consequence, the freeze-out dynamics is controlled by a thermal bath with a number of light degrees of freedom at the freeze-out temperature $g_*^{\text{f.o.}}\equiv g_*(T_{\text{f.o.}})$ possibly different than the SM one.

\subsection{Hard cross-section and Sommerfeld enhancement}\label{sec:hardLOandSE}

\begin{figure}[htp!]
    \centering
    \includegraphics[scale=0.55]{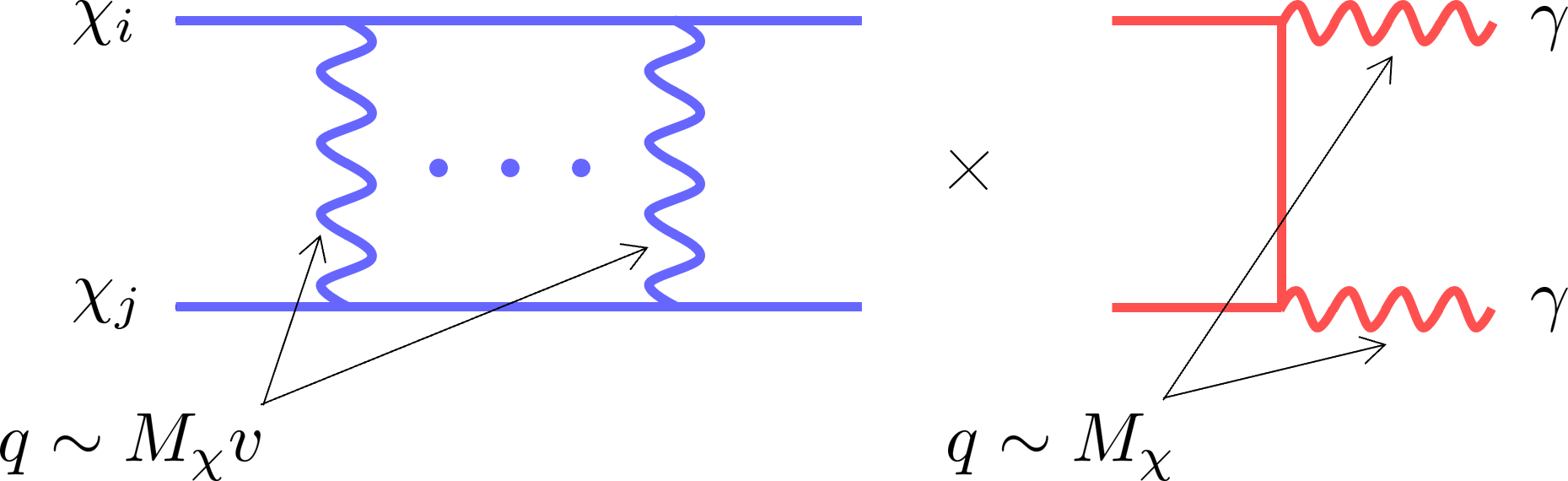}
    \caption{Sketch of the annihilation rate process. The {\bf red} diagram shows the hard annihilation cross-section where the particles in the final state carry most of the DM energy. The {\bf blue} diagram includes soft exchanges with momentum $q\sim M_\chi v$ which get resummed by the non-relativistic potential.}
    \label{fig:LOsom}
\end{figure}
In the non-relativistic limit the dynamics of annihilation is captured by the Schr\"{o}dinger equation for the system of the pair of annihilating particles $\phi_{j,i}(r)$. Here $j=s$ selects a given total angular momentum channel at $l=0$ while the index $i$ stands for other possible internal degrees of freedom. The Schr\"{o}dinger equation reads \cite{Blum:2016nrz}
\begin{equation}
\label{eq:Schrodinger}
-\frac{\nabla^2\phi_{j,i}}{M_\chi}+\left[V_{i}^{\text{LO}}-i \frac{\sigma_{j,i}^h \vrel}{2}\delta^3(\vec{r})\right]\phi_{j,i} = \frac{M_\chi\vrel^2}{4}\phi_{j,i}\, ,
\end{equation}
the imaginary part of the potential in the squared brackets is related through the optical theorem to the ``hard'' annihilation cross-sections $\sigma_{j,i}^h$ which describes the processes whose decay products carry most of the DM energy as shown in Fig.~\ref{fig:LOsom}. Upon projecting into radial wave functions, the kinetic term will also generate the usual centrifugal barrier $1/r^2$.  

At small velocities, $\vrel\ll \alpha$, the dynamics of DM is affected by the long-range interactions encoded in the non-relativistic potentials $V_{i}^{\text{LO}}$. At leading order in $\alpha$ and $\vrel$, the potential takes the standard Coulomb form 
\begin{equation}
\vlo_{i}(r)= \lambda_i\frac{\alpha}{r}\ ,\label{sec:potentialLO}
\end{equation}
where $\lambda_i$ is a channel-dependent number (which we assume to be spin independent) that can be negative (positive) for attractive (repulsive) potentials.\footnote{We refer to Ref.~\cite{Bellazzini:2013foa} for a complete classification of the possible Coulomb interaction from the effective field theory perspective. We also assumed the mass $M_\chi$ to be the same for every internal degree of freedom $i$. This is easily generalized in the case where mass-splittings among the different internal degrees of freedom play an important role (see for instance Ref.~\cite{Bottaro:2022one}).} 

The full, non-perturbative annihilation cross-section can be computed by solving Eq.~\eqref{eq:Schrodinger}. The contribution of the modified wave function to the annihilation cross-section can be read from the divergence of the probability current $\vec{j}_{j,i}(\vec{r})=\frac{2}{M_\chi} \Im[\chis^\dagger\vec{\nabla}\chis]$
\begin{equation}
\label{eqA:xsec}
\sigma_{j,i}\vrel = -\int\mathrm{d}^3r\vec{\nabla}\cdot\vec{j}_{j,i}(\vec{r})\simeq\shard_{j,i}\vrel |\phi_i(0)|^2\ ,
\end{equation}
where $S_E^i\equiv |\phi_i(0)|^2$ is the so-called Sommerfeld enhancement (SE) and $\phi_i(x)$ is the spin-independent solution of the Schr\"{o}dinger equation  in Eq.~\eqref{eq:Schrodinger} where the contribution from the hard process is neglected.\footnote{Going beyond the approximation in Eq.~\eqref{eqA:xsec} including the effect of the hard process on the wave function introduces corrections to the SE of order $\mathcal{O}(\alpha^3)$ which are negligible for DM freeze-out (see Ref.~\cite{Blum:2016nrz} for a discussion of the importance of these corrections in the context of indirect detection).} The boundary conditions for $\phi_{i}$ are requiring regularity at the origin and to recover the asymptotic scattering waves away from the potential. 

A nice way of writing the SE is to introduce the reduced radial wave function for $l=0$, $\chi_{j,i}(r)= r\phi_{j,i}(r)$ and define the dimensionless variable $x=r M_\chi$ and the rescaled potential $V_i^{\text{LO}}=M_\chi \mathcal{V}^{\text{LO}}_i$ so that Eq.~\eqref{eq:Schrodinger} becomes 
\begin{equation}
-\chi''_{j,i}+\left[\mathcal{V}_i^{\text{LO}}-iM_\chi^2\frac{\sigma^h_{j,i} v}{2}\delta^3(\vec{x})\right]\chi_{j,i}=\frac{v^2}{4}\chi_{j,i}\ .\label{eq:Sdimless}
\end{equation}
Choosing carefully the boundary conditions we can get equivalent expressions of the SE in terms of the reduced wave function. For instance taking $\chi(0)\propto x$ and $\chi_i(\infty)=\sin(x v/2+\delta_0^i)$ we get $S_E^i=\vert 2\chi_i'(0)/v\vert^2$. Alternatively, imposing $\chi(0)=xv/2$ we get $\chi_i(\infty)=\sin(x v/2+\delta_0^i)/\sqrt{S_E}$. This latter formulation will be useful in Sec.~\ref{sec:UV}. 

Another simple way to compute the SE is to rewrite Eq.~\eqref{eq:Schrodinger} as a first order differential equation for $h_{j,i}(r)=\chi_{j,i}'(r)/\chi_{j,i}(r)$
\begin{equation}
  \frac{1}{M_\chi}(h'_{j,i}+h^2_{j,i})-\left[V_i^{\text{LO}}-i \frac{\sigma_{j,i}^h \vrel}{2}\delta^3(\vec{r})\right]= -\frac{M_\chi v^2}{4}\, ,
\end{equation}
with boundary condition $h(x\rightarrow\infty)=i \mdm v/2$. In terms of the new variable the SE in Eq.~\eqref{eqA:xsec} can be written as $S_E^i=2\Im [h_i(0)]/(\mdm v)$ under the same approximation of Eq.~\eqref{eqA:xsec}. We will use this formulation in Sec.~\ref{sub:IR}. 

\begin{figure*}[htp!]
    \centering
    \includegraphics[scale=0.5]{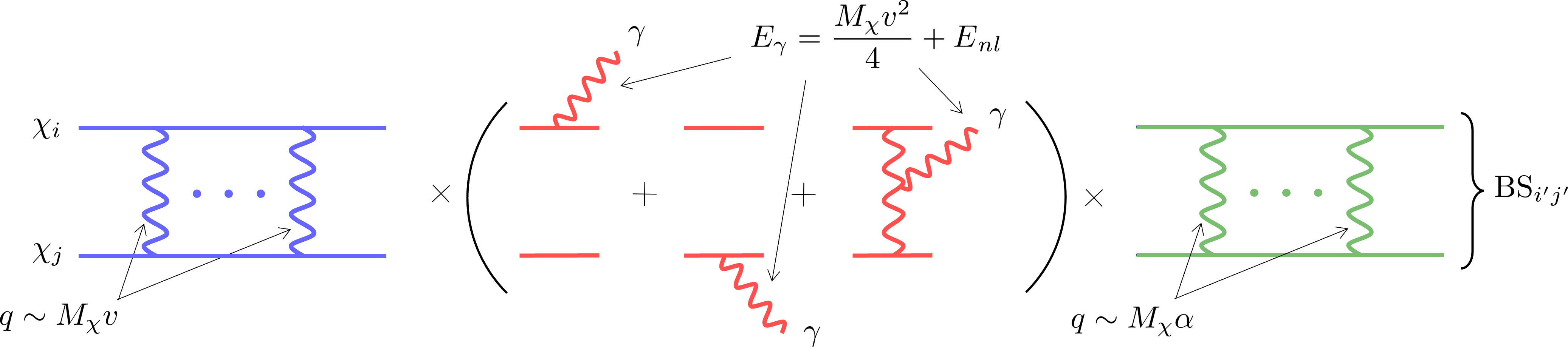}
    \caption{Sketch of the bound state formation process. The {\bf red} diagram shows the fundamental emission process, where a vector with energy given approximately by the binding energy of the BS is emitted from the DM line (dipole) or from the internal vector line (non-abelian). The {\bf blue (green)} ladder diagram includes soft photon exchanges with momentum $q\sim M_\chi v$ ($q\sim M_\chi \alpha$) which get resummed by the non-relativistic potential at LO into the scattering (bound state) wave function.}
    \label{fig:BSF}
\end{figure*}

For a Coulomb potential, the $S_E$ takes the simple analytic form
\begin{equation}
\label{eq:som_en}
S_E^i\vert_{\text{LO}} = \frac{2\pi\lambda_i\alpha}{\vrel}\frac{1}{1-e^{-\frac{2\pi\lambda_i\alpha}{\vrel}}}\ ,
\end{equation} 
which strongly enhances (or suppresses) the $l=0$ annihilation cross-section when $\vrel<\lambda_i\alpha$. The annihilation cross-section for a single $j$ wave can then be parametrically written at leading order as
\begin{equation}
  \sigma^{j}_{\text{ann}}v=\sum_i\sigma_{j,i}v= \frac{2\pi^2\alpha^3}{vM_\chi^2} c_\text{ann}^j\vert_{\text{LO}}\ ,
\end{equation}
where we defined $c_\text{ann}^j\vert_{\text{LO}} =\sum_{i}c_{j,i}^h\lambda_i$ and $c_{j,i}^h$ are $\mathcal{O}(1)$ coefficients encoding the leading order contribution of the hard cross-section to the different channels and $\lambda_i$ are defined in Eq.~\eqref{sec:potentialLO}.

\subsection{Bound State Formation}\label{subsec:BSFLO}

In the non-relativistic regime typical of freeze-out, long-range interactions lead to a significant rate of BSF. Since these bound states are not stable, their formation and later decay helps in depleting the DM abundance and thus affects the prediction of the DM mass. 

At leading order in $\alpha$ and $\vrel$, BSF occurs via emission of a single gauge boson through the process $\chi_{1,i}+\chi_{2,i'}\rightarrow \text{BS}_{ff'}+V^a$ described by a generalized dipole Hamiltonian derived in Ref.~\cite{Beneke:1999zr,Manohar:2000kr,Harz:2018csl}. We can write the final BSF cross-section as
\begin{equation}
    \sigma_{\text{BS}}^{i_si_b}\vrel=S_{\text{BS}}^{i_si_b}\left( S_E^{i_s}\frac{\pi\alpha^2}{M_\chi^2}\right)\, ,\label{eq:schematically}
\end{equation}
where $S_E^{i_s}\simeq 2\pi\lambda_{i_s}\alpha/v$ is the SE corresponding to the incoming scattering state, while $S^{i_si_b}_{\text{BS}}$ encodes the non-trivial coupling and velocity dependence coming from the overlap integral between the incoming scattering state $i_s$ and the outcoming bound state. We leave a detailed discussion of the overlap integrals to Appendix~\ref{sec:BSFNLO}.

\subsection{Unitarity bound at leading order}\label{subsec:uniLO}
As a direct consequence of the unitarity of the $S$-matrix, the 2-2 annihilation cross-section in a given partial wave with total angular momentum $j$ is bounded from above by $\sigma_{j}\leq \pi(2j+1)/p^2$, where $p$ is the initial momentum of the annihilating particles. In the non-relativistic limit, $p^2= M_\chi^2v^2/4$ and the perturbative unitarity bound can be written as 
\begin{equation}
\sigma_{j} v\leq \frac{4\pi(2j+1)}{M_\chi^2 v}\ .\label{eq:PUB}
\end{equation}
This inequality does not require the existence of free asymptotic states and it is also satisfied for scattering processes in a Coulomb potential~\cite{Landau:1991wop}. 

The annihilation cross-section at leading order and in the limit $v\ll\alpha$ can be written as 
\begin{equation}
\sigma_{j}\vrel\vert_{\text{LO}}\simeq\frac{2\pi^2\alpha^3}{vM_\chi^2}\left[c_\text{ann}^j\vert_{\text{LO}}+c_{\text{BSF}}^j\vert_{\text{LO}}\right]\ \ , \label{eq:full}
\end{equation}
where the coefficients $c_\text{ann}^j\vert_{\text{LO}}$ and $c_{\text{BSF}}^j\vert_{\text{LO}}$ encode model dependent $\mathcal{O}(1)$ numbers controlling the annihilation cross-section and the BSF respectively. In particular, $c_\text{BSF}^j$ can be obtained from Eq.~\eqref{eq:schematically} by projecting onto a state with total angular momentum $j$. 

The maximal coupling allowed by the unitarity bound at leading order is 
\begin{equation}
\alpha^{\text{max}}_{\text{LO}}=\text{Min}_j\left[\frac{2(2j+1)}{(c_\text{ann}^j\vert_{\text{LO}}+c_{\text{BSF}}^j\vert_{\text{LO}})\pi}\right]^{1/3}\ .\label{eq:PUBLO}
\end{equation}
It is interesting to notice that the SE in the non-relativistic limit reduces the maximally allowed coupling by roughly one order of magnitude compared to the usual bound from the relativistic power counting.   

To a good approximation, the solution of the DM Boltzmann equation is equivalent to requiring the freeze-out yield to be $Y_\chi^{\text{f.o}}\simeq \frac{H}{s\sum_j\langle\sigma_j v\rangle}$, where $s$ is the entropy density. Substituting $\alpha^{\text{max}}_{
\text{LO}}$ in the freeze-out condition and requiring $\chi$ to account for the DM relic density today we get the maximal DM mass allowed by the perturbative unitarity bound 

\begin{widetext}
\begin{equation}
M_\text{max}\simeq 131.7\text{ TeV}\left(\alpha^{\text{max}}_{
\text{LO}}\right)^{3/2}\left(\sum_j (c_\text{ann}^j\vert_{\text{LO}}+c_{\text{BSF}}^j\vert_{\text{LO}})\right)^{1/2}\left(\frac{g_*^{\text{f.o.}}}{x_{\text{f.o}}}\right)^{1/4}\ ,\label{eq:mmax}
\end{equation} 
\end{widetext}

where we substituted the high temperature value of the SM degrees of freedom $g_*^{\text{SM}}\simeq106.75$ and $x_{\text{f.o}}\equiv M_\chi/T_{\text{f.o.}}\simeq 29$ (see Ref.~\cite{Steigman:2012nb} for details about the dependence of $x_{\text{f.o}}$ on the model parameters). 

If the annihilation is dominated by a single $j$-wave then Eq.~\eqref{eq:mmax} simplifies and the mass upper bound becomes independent on the model as derived in Ref.~\cite{Griest:1989wd}. For example if the $j=0$ wave dominates we get $M_{\text{max}}^{J=0}=137\text{ TeV}$ for a non-self conjugate DM particle (in the self-conjugate case the maximal DM mass will be heavier by a factor of $\sqrt{2}$).

In practice, the dominance of a single $j$-wave is an unrealistic assumption in most of the freeze-out scenarios. The main reason is that BSF tends to equally distribute the cross-section in the lower $j$ channels. As a consequence, accounting for BSF generically makes $M_{\text{max}}$ larger with respect to the naive estimate because the contribution from BSF in Eq.~\eqref{eq:mmax} overcomes the tightening of the perturbative unitarity bound on $\alpha$ in Eq.~\eqref{eq:PUBLO}.\footnote{In previous works on the subject \cite{vonHarling:2014kha,Smirnov:2019ngs} the upper bound on the DM mass is typically the one obtained assuming the dominance of the $j=0$ partial wave (see however the discussion in Ref.~\cite{Baldes:2017gzw}). Even for this simplified setup in Ref.~\cite{vonHarling:2014kha} no decomposition in partial waves of the \textit{total} angular momentum was made, and the total annihilation cross-section, including BSF, was compared to the PUB on the $j=0$ wave, thus significantly underestimating the upper bound on the coupling. This error was later amended in Ref.~\cite{Baldes:2017gzw} which agrees with the number derived here.}

\begin{figure*}[htp!]
    \centering
    \includegraphics[scale=0.45]{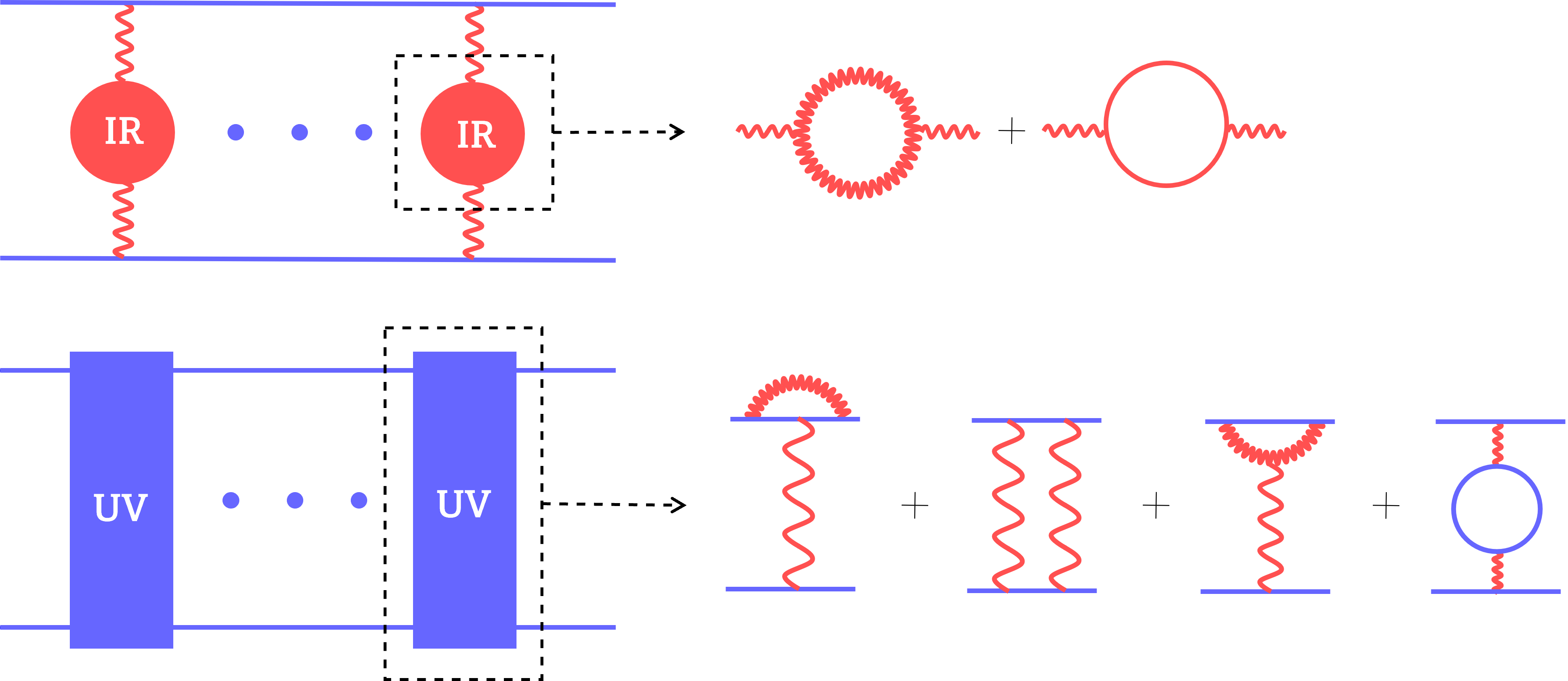}
    \caption{Sketchy representation of the IR contributions ({\bf top}) and UV contributions ({\bf bottom}) to the non-relativistic potentials at NLO. {\bf Red} lines corresponds to IR degrees of freedom: fermions ({\bf solid}) or gauge bosons ({\bf wiggly}). {\bf Blue} lines are instead heavy DM lines which we take to be scalar in this paper.}
    \label{fig:NLOPOT}
\end{figure*}
\section{NLO corrections}\label{sec:NLO}
\label{sec:DQEDheavy}
We investigate the NLO correction to the non-relativistic potential. These are expected to be the dominant NLO corrections near the unitarity bound at non-relativistic velocities $v\ll\alpha$ typical during freeze-out.\footnote{In Sec.~\ref{sec:DQED} we focused on $s$-wave annihilation processes ($\vec{L}=0$) claiming that higher $L$ waves would be velocity suppressed. While this is certainly true for the hard cross-section, including SE for a general wave gives $S_E^j\times\sigma_j^h v\sim \alpha^{2l}\times \sigma_0^{h} v\ ,$~\cite{Cassel:2009wt}. As a consequence $p$-wave contributions to the hard annihilation process are only $\alpha^2$ suppressed with respect to the $j=0$ channel and should be included approaching the unitarity bound. NLO correction will also affect the hard annihilation cross-section but are expected to be subleading with respect to the corrections to the non-relativistic potentials for $v\ll\alpha$.}   

NLO corrections arise from both IR and UV dynamics. IR corrections result from loops involving only light degrees of freedom and affect both the long- and the short-range behavior of the Coulomb potential. In contrast, UV corrections modify only the short-range potential, originating from diagrams where the heavy DM field $\chi$ also flows in the loops. Both contributions are schematically shown in Fig. \ref{fig:NLOPOT}. 

The dominant IR corrections can be approximated by the including the running coupling $\alpha(r)$ in the leading order Coulomb potential~\cite{Beneke:2019qaa,Beneke:2020vff,Urban:2021cdu} (see also a Ref.~\cite{Harz:2018csl,Harz:2019rro,Binder:2020efn} for discussions about the role of IR NLO corrections)

\begin{equation}
\label{eq:NLO_IR}
    V_{\rm NLO}^{\rm IR}=-\frac{\alpha(r)}{r}\ , \quad \alpha(r)=\frac{\alpha}{1+\frac{\alpha}{2\pi}b^{\rm IR}\log(M_\chi r)}\ ,
\end{equation}
where $b^{\rm IR}$ is the beta-function coefficient due to the states lighter than the DM mass running in the loops and we defined $\alpha$ to be the fine structure constant at the DM threshold $r=1/M_\chi$.

UV corrections can be systematically captured via one-loop matching to the DM potential in Non-Relativistic QFT \cite{Beneke:2013jia}. These thresholds are organized in powers of $1/M_\chi$ whose general form in Fourier space was derived in Ref.~\cite{Beneke:2013jia} up to order $\mathcal{O}(1/M_\chi^2)$. In position space, the NLO potential reads
\begin{equation}
    \Delta V_{\rm NLO}^{\rm UV}=-\frac{\alpha}{r}\left[\frac{V_2}{M_\chi r}+\frac{V_3}{M_\chi^2r^2}\right]+\frac{V_D\alpha}{M_\chi^2}\delta^3(\vec{r})\ .\label{eq:NLO_UV}
\end{equation}
where $V_{2}$, $V_3$, and $V_D$ are dimensionless coefficients that may depend on the annihilation channel (including the spin). $V_3$ and $V_D$ contain in general spin-dependent terms like spin-spin and spin-orbit interactions. These terms induce transitions between states with different $l$ and $s$. In general, computing SE and BSF would require solving a system of coupled Schr\"{o}dinger equations. In the following, we  consider the simpler case of scalar DM, where these terms are zero. We provide more details on the general structure of the NLO potential in Appendix \ref{app:fourier}, see also Ref.~\cite{Beneke:2020vff} where the case of fermionic DM in a triplet representation of $SU(2)$ is discussed.

Using the DM De Broglie wavelength, $r\sim\lambda_B\equiv 1/(M_\chi v)$, we estimate the NLO corrections as $\Delta V_\text{NLO}\sim [V_2 v+ (V_3+V_D) v^2]V_\text{LO}$ where $V_{\text{LO}}\sim \alpha/\lambda_B= \alpha M_\chi v$ is the scaling of the Coulomb potential in Eq.~\eqref{sec:potentialLO}. This estimate indicates that the inverse square potential dominates over the other NLO corrections for $v< V_2^i/(V_3^i+V_D^i)$. The leading relativistic correction to the kinetic energy $M_\chi^3\Delta E_{\text{kin}}=p^4/8 $ scales as $\Delta E_{\text{kin}}\sim M_\chi v^4$ and are negligible with respect to the correction to the potential as long as $v< (V_2 \alpha)^{1/2}$ and $v< (V_3+V_D) \alpha$ respectively. This shows that at low enough velocities, the leading NLO corrections can be encoded solely in the modifications of the potential in the Schr\"{o}dinger equation. 

We now present a systematic procedure to account for both IR and UV NLO corrections in the SE and in BSF. 

\subsection{IR contributions at NLO}\label{sub:IR}

The IR behavior of the NLO potential crucially depends on the sign of the IR beta-function coefficient $b_{\rm IR}$. Here we focus on IR free theories with $b_{\rm IR}>0$. In such cases, the NLO potential exhibits a Landau pole at a position $ r_{*}M_\chi=\exp(-\frac{2\pi}{b_{\rm IR}\alpha})$. The presence of this Landau pole prevents fixing boundary conditions at distances arbitrarily close to the origin. 

The boundary condition is instead set at a distance $r_p$, defined as $2\Im [h(r_p)]/\mdm\vrel=(1-p) S_E$, where $p$ is an arbitrary parameter that controls the theoretical accuracy of the approximation. By inspecting the Schr\"{o}dinger equation, we can estimate $r_p$ as the point where the SE reaches its plateau: $p/r_p \approx2(1+\pi)M_\chi\alpha$. The value of $r_p$ is independent of velocity because, at small distances, the potential energy dominates over the kinetic energy. The larger $p$, the further we are from SE saturation, leading to larger theoretical errors. 

As long as $ r_{p}>r_*$, we can solve the Schr\"{o}dinger equation with the resummed potential in Eq.~\eqref{eq:NLO_IR}, and compute the SE. For $\alpha_{\text{LO}}^{\text{max}}$ defined in Eq.~\eqref{eq:PUBLO} $r_*$ is the largest: $r_*(\alpha_{\text{LO}}^{\text{max}})=r_*\vert_{\text{max}}$. To minimize the theoretical error at the perturbative unitarity bound, we fix $p$ by setting $r_{\bar{p}}=r_*\vert_{\text{max}}$, thus determining $p=\bar{p}$. This ensures the calculability of the SE up to the perturbative unitarity bound. We then estimate the theoretical error due to IR NLO effects by taking the difference of the SE between $r_{\bar{p}}$ and $2 r_{\bar{p}}$ for any coupling: $\Delta S^{\text{IR}}_E= S_E(r_{\bar{p}})- S_E(2r_{\bar{p}})$. In Sec.~\ref{sec:darkQED} we will apply this recipe to a simple toy model with abelian gauge interactions. 

We now briefly comment on the impact of IR NLO corrections on BSF. The bound state wave functions and binding energies can be computed at NLO by using the Coulombian bound states as an unperturbed basis and diagonalizing the matrix elements of the Hamiltonian $\mathcal{H}_{\text{NLO}}=\frac{p^2}{M_\chi}+V_{\text{NLO}}^{\text{IR}}(r)$. The matrix elements are given by $\mathcal{H}_{ij}^l\equiv\langle\Psi^C_{il}|\mathcal{H}|\Psi^C_{jl}\rangle$, where $|\Psi^C_{il}\rangle$ represents the Coulombian bound state with principal quantum number $i$ and angular momentum $l$. The bound state wave function and its binding energy are mostly sensitive to scales larger than the Bohr radius, $r\gtrsim\frac{1}{M_\chi\alpha}$. For this reason, the bound state dynamics is insensitive to NLO corrections in theories with $\beta_{\text{IR}}>0$, since the deviations from the Coulomb behavior occur at short scales. Conversely, we expect large NLO corrections to the bound state dynamics in UV-free theories with $\beta_{\text{IR}}<0$. We leave a detailed study of this case for the future.

For $\beta_{\text{IR}}>0$, the BSF cross-section is mostly affected by NLO modifications of the scattering wave function. These corrections particularly affect the formation of $p$-wave bound states from a $s$-wave initial scattering states, as scattering waves with angular momentum $l>0$ are screened from short-scale NLO corrections by the centrifugal potential. More details are given in Appendix~\ref{sec:BSFNLO}.

\subsection{UV contributions at NLO}\label{sec:UV}

The inclusion of the UV NLO corrections in Eq.~\eqref{eq:NLO_UV} makes the Hamiltonian no longer bounded from below. Equivalently,  the Schr\"{o}dinger equation with $V_{\text{NLO}}^{\text{UV}}=V_{\text{LO}}+\Delta V_{\text{NLO}}^{\text{UV}}$ cannot be solved with normalizable solutions with boundary conditions at the origin. This poses the challenge of defining the SE in the presence of UV singular potential. 

To address this, we follow the approach of Ref.~\cite{Bellazzini:2013foa} (see also Ref.~\cite{Lepage:1997cs,Kaplan:1996nv} for similar techniques applied to nuclear physics). We regularize the full potential close to the origin with a well potential at distances $r<r_{\rm cut}\equiv\xcut/M_\chi$. Using the dimensionless variable $x=r M_\chi$ we can define a dimensionless regularized potential 
\begin{equation}\label{eq:reg_pot}
\mathcal{V}_{\text{reg}}(v,x)=\left\{
\begin{aligned}
&-\vcut(
v,x_{\text{cut}}),\quad x<\xcut\\
&\mathcal{V}_{\text{NLO}}^{\text{UV}}(x),\qquad\quad\ \ x>\xcut
\end{aligned}\right.\ ,
\end{equation}
where $V_{\text{NLO}}^{\text{UV}}=M_\chi \mathcal{V}_{\text{NLO}}^{\text{UV}}$.
The Schr\"{o}dinger equation in Eq.~\eqref{eq:Sdimless} can be written by substituting the regularized potential ($\mathcal{V}_{\text{LO}}\to \mathcal{V}_{\text{reg}}$). The kinetic term is also modified by an arbitrary wave function renormalization, $\chi''\to Z_\chi \chi''$, for $x<x_{\text{cut}}$. In general, both the depth of the potential well $-\vcut$ and the wave function renormalization $Z_\chi$ depend on $x_{\text{cut}}$ and $v$, and they must be fixed with appropriate renormalization conditions in order for the SE to be well defined. 

Since the relativistic UV theory is calculable, the scattering phase can be explicitly computed and matched to the non-relativistic EFT as an input (see Ref.~\cite{Agrawal:2020lea,Parikh:2020ggm} for a similar discussion in the context of scattering processes). At distances $x<\xcut$, the $s$-wave scattering phase can be computed for a generic central potential in the Born approximation 
\begin{equation}
\label{eq:prateek}
\sin\bar{\delta}_0^{\text{NLO}}=-\frac{\vrel}{2}\int_0^{x_{\text{cut}}}\mathrm{d}x x^2 j_0^2\left(\frac{vx}{2}\right)\mathcal{V}_{\text{NLO}}(x)\ ,
\end{equation}
where $j_0(x)$ is the regular, spherical Bessel function and $\mathcal{V}_{\text{NLO}}$ is the full NLO potential. 

The UV scattering wave function can then be written as $\sin(v x/2+\bar{\delta}_0)$ and it is then matched at $x=x_{\text{cut}}$ to the solution of the Schr\"{o}dinger equation for $x<x_{\text{cut}}$ inside the potential well. By matching the logarithmic derivatives we get 
\begin{equation}
\tan\left(\frac{v\xcut}{2}+\bar{\delta}_0^{\text{NLO}}\right)=\frac{v}{2\Phi_v}\tan\left(\xcut\Phi_v\right)\ ,\label{eq:matching}
\end{equation}
where $\Phi_v\equiv\left(\frac{4\vcut+v^2}{4Z_\chi}\right)^{1/2}$. At fixed $x_{\text{cut}}=1$ and $v$, this matching condition determines $\Phi_v$ in terms of $\bar{\delta}_0^{\text{NLO}}$. The solution inside the potential well is then
\begin{equation}
\chi_v^{\text{in}}(x<\xcut)=\frac{v}{2\Phi_v}\sin(\Phi_v x)\label{eq:insidethewell}
\end{equation}
and can be matched at $x=x_{\text{cut}}$ to $\chi^{\text{out}}_v(x)$, the solution to the  Schr\"{o}dinger equation for $x>x_{\text{cut}}$ with the potential in Eq.~\eqref{eq:reg_pot}. However, this boundary condition alone does not fully determine the SE, as the asymptotic behavior of $\chi^{\text{out}}_v(x)$ also depends on the wave function renormalization $Z_\chi$
\begin{equation}
\lim_{x\to\infty}\chi_v^{\text{out}}(x)=\frac{\sin\left(\frac{xv}{2}+\delta\right)}{\sqrt{S_E(v)Z_\chi}}\ ,\label{eq:extractingSE}
\end{equation}    
where the phase $\delta$ depends on the long-range dynamics and has to be distinguished from the short-distance one in Eq.~\eqref{eq:prateek}.

To extract the Sommerfeld enhancement from \eqref{eq:extractingSE}, we need an additional condition to determine the wave function renormalization, $Z_\chi$. Here we follow the procedure first proposed in Ref.~\cite{Bellazzini:2013foa}, where $Z_\chi$ is obtained by exploiting the fact that the SE must approach 1 at large velocities. The physical intuition behind this limit  can be understood by noticing that, as the velocity increases, the variation of the potential within the DM de Broglie wavelength $\lambda_B=1/M_\chi v$ becomes negligible. This reduced variation means the potential has less impact on deforming the wave functions of DM scattering states, resulting in a diminished Sommerfeld enhancement. As velocity approaches 1, we expect the SE to approach 1 as well. This is obviously only a mathematical, formal limit, since the non-relativistic effective theory is no longer valid as we approach the speed of light, and the correct dynamics cannot be described by a Schr\"{o}dinger equation. Besides, in practice the Sommerfeld enhancement is not exactly 1 but $S_E(v\rightarrow 1)=1+\mathcal{O}(\alpha)$, as can be explicitly checked in the case of the Coulomb potential \eqref{eq:som_en}. The $S_E\rightarrow 1$ only as $v\rightarrow \infty$, because in this unphysical limit, the de Broglie wavelength becomes zero, causing the potential’s effect to vanish completely. In the following, we will employ this purely formal infinite velocity limit to remove the spurious $\mathcal{O}(\alpha)$ corrections to $Z_\chi$ coming from the Sommerfeld enhancement, as we show below. 

Following \cite{Bellazzini:2013foa}, to fix $Z_\chi$ we set up a different scattering problem where two DM particles scatter with relative velocity $w$ inside the potential defined in Eq.~\eqref{eq:reg_pot}. The  Schr\"{o}dinger equation for $x<x_{\text{cut}}$ reads 
\begin{equation}
-Z_\chi(\chi^{\text{in}}_w)''-\mathcal{V}_{\text{cut}}(v,\xcut)\chi_w^{\text{in}}=\frac{w^2}{4}\chi_w^{\text{in}} \ .\label{eq:zchi}
\end{equation}
Notice that $\mathcal{V}_{\text{cut}}(v,\xcut)$ is determined by the original velocity $v$ at which we are computing the Sommerfeld enhancement. The solution to \eqref{eq:zchi} is given by
\begin{equation}\label{eq:insidethewellZchi}
\begin{split}
&\chi_{w}^{\text{in}}(x<\xcut)=\frac{w}{2\Phi_{w}}\sin(\Phi_{w} x)\ ,\\
&\Phi_{w}=\left(\frac{4\vcut(v,\xcut)+w^2}{4Z_\chi}\right)^{1/2} \ ,
\end{split}
\end{equation}
which fixes the initial conditions at $x=\xcut$. Solving the Schr\"{o}dinger equation outside the potential well and requiring  $S_E(w\gg 1)=1$ yields
\begin{equation}
 \lim_{x\to\infty} \chi_{w}^{\text{out}}(x)\stackrel[w\rightarrow\infty]{}{\simeq}\frac{\sin\left(\frac{w x}{2}+\delta\right)}{\sqrt{Z_\chi}}\ ,\label{eq:bc2}
\end{equation}
which is equivalent to Eq.~\eqref{eq:extractingSE} but with the Sommerfeld factor now set to one.
Combining \eqref{eq:insidethewellZchi} and \eqref{eq:bc2} we can then extract $Z_\chi$. Taking the $w\rightarrow\infty$ limit corresponds to the system’s kinetic energy becoming infinitely large relative to the potential energy. In this case, the average potential energy within a region of size $\lambda_B=1/(M_{\chi}w)$, for large enough $w$, is simply $\langle \mathcal{V}_{\rm reg}\rangle = \mathcal{V}_{\rm cut}$, which quickly becomes negligible  compared to the kinetic energy $M_\chi w^2/4$. Taking the formal limit of $w\rightarrow\infty$ is then equivalent to switching off the regularized potential, justifying $S_E(w\rightarrow\infty)=1$ in Eq.~\eqref{eq:bc2}.\footnote{It appears that Ref.~\cite{Bellazzini:2013foa} imposes a similar renormalization condition than Eq.~\eqref{eq:bc2} but requiring $S_E(w\rightarrow 1)=1$. In Appendix \ref{sec:coulomb} we show that this condition applied to the simple case of the Coulomb potential  leads to a significantly worse agreement with the exact result in Eq.~\eqref{eq:som_en}.}


The matching condition in Eq.~\eqref{eq:matching} admits different branches of solutions for $\Phi_v^i$, each leaving the l.h.s of Eq.~\eqref{eq:matching} unchanged~\cite{Beane:2000wh}. However, only the solution in the first branch, defined by $0<\Phi_v<\pi/2\xcut$, leads to a well-defined SE given the boundary conditions in Eq.~\eqref{eq:matching} and Eq.~\eqref{eq:bc2}.

This can be seen by considering the case of a very short-range potential with depth $\vcut$, which vanishes immediately outside the potential well $x>x_{\text{cut}}$. This setup effectively captures the physics in the extremely weakly coupled regime. In this case, the wave function outside the well becomes a simple plane wave. Following the UV matching procedure, we find the SE
\begin{equation}
    S_E=\frac{1}{Z_\chi\cos^2(\Phi_v\xcut)\left[1+\frac{v^2}{4}\frac{\tan^2(\Phi_v\xcut)}{\Phi_v^2}\right]}\, .\label{eq:branchdep}
\end{equation} 
Requiring $S_E(v\gg1)=1$ as in Eq.~\eqref{eq:bc2} fixes $Z_\chi=1$. 

Eq.~\eqref{eq:branchdep} shows explicitly how the SE depends on the branch through the $1/\cos^2(\Phi_v\xcut)$ factor. Defining the $i$-th branch as the one with $(2i-1)\pi/2\xcut<\Phi_v<(2i+1)\pi/2\xcut$, with $i$ integer, the SE increases by moving to higher branches because the $\cos^2(\Phi_v\xcut)$ decreases. This behavior is inconsistent with the weak coupling limit ($\mathcal{V}_{\rm cut}\ll v^2/4$ in this simple setup) where we expect to recover $S_E\to 1$. This limit is realized in the first branch, which we select as the physical one.\footnote{One might wonder if this argument depends on the simplicity of the wave function in the simple example of the short range potential well. However, the same argument can be constructed starting from the Coulomb potential $V(r)=\alpha/r$ and using both the regular and the irregular hypergeometric function to perform the UV matching~\cite{abramowitz+stegun}. We checked that in the limit $\alpha\ll v\ll 1$ this exercise leads to an expression very similar to Eq.~\eqref{eq:branchdep} where the 1 at the numerator is replaced by the SE at LO in Eq.~\eqref{eq:som_en}.} 

\begin{figure*}[htp!]
\centering
\includegraphics[width=0.485\textwidth]{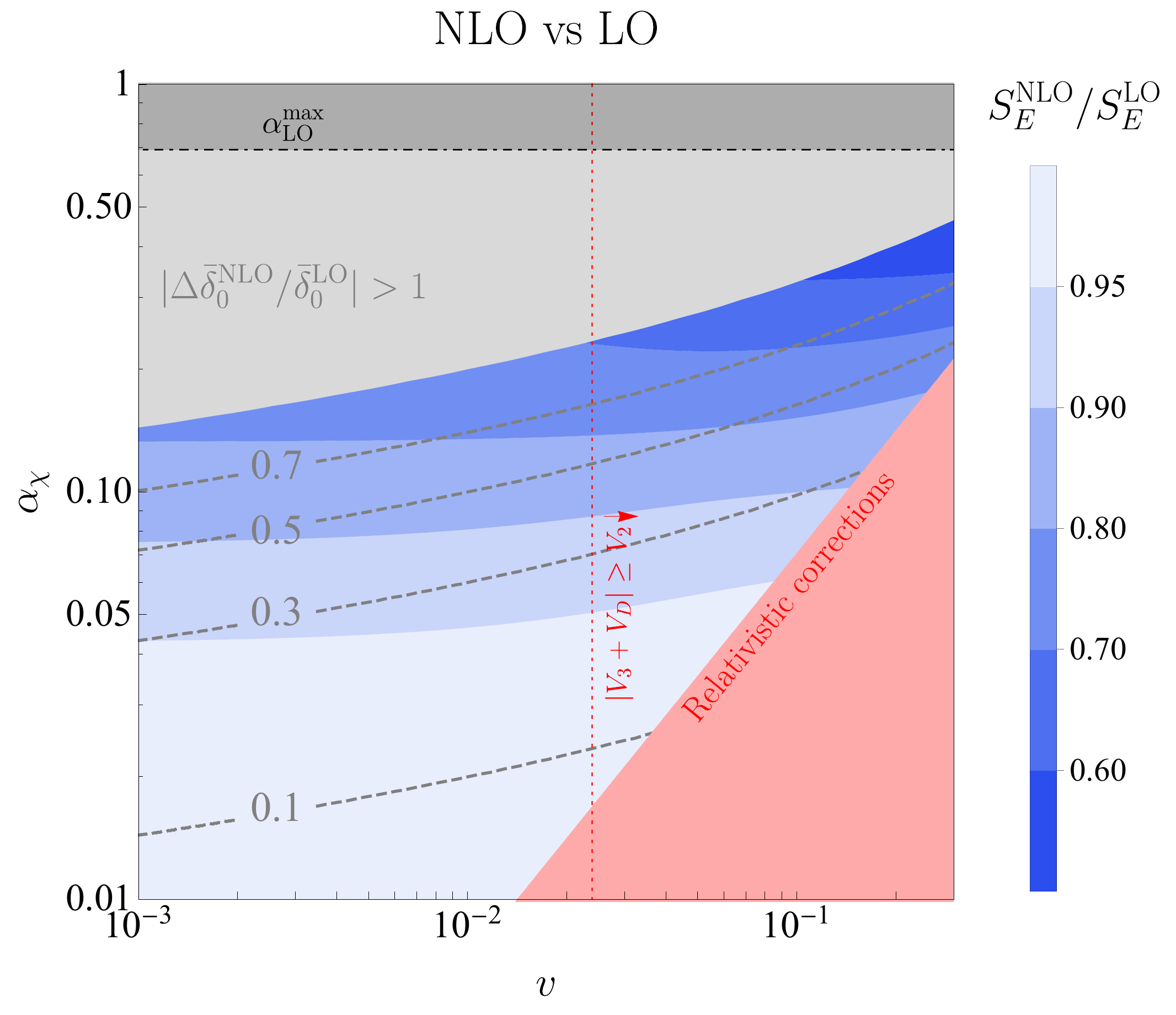}\hfill
         \includegraphics[width=0.5\textwidth]{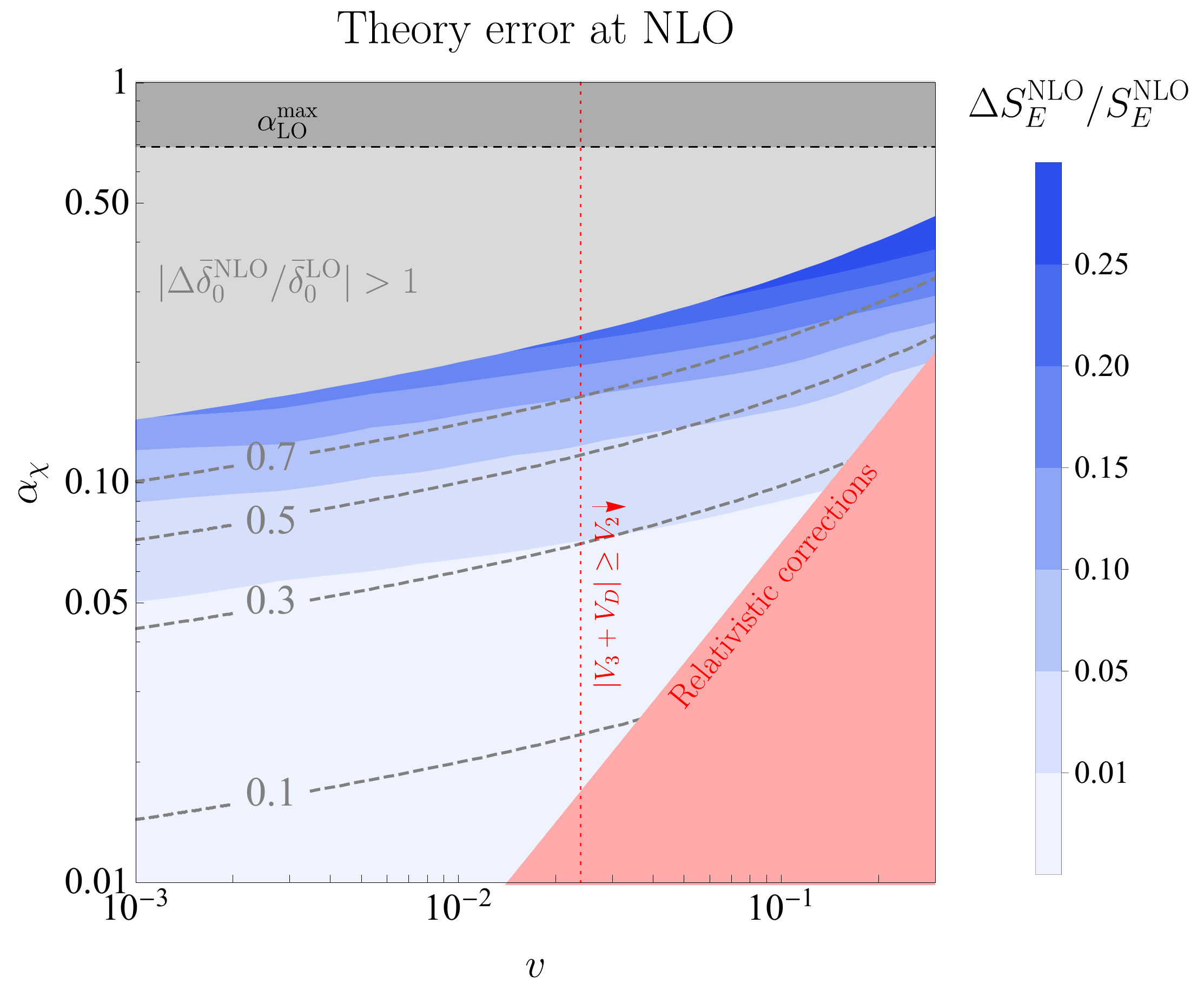}
\caption{Behavior of the UV NLO corrections described in Sec.~\ref{sec:UV} with $x_{\text{cut}}=1$. The red lines exhibit the power counting on the relativistic corrections discussed below Eq.~\eqref{eq:NLO_UV}. On the right of the {\bf red} dashed line, the NLO potential cannot be approximated with a $1/r^2$ correction to the Coulomb potential while on the {\bf red} shaded region the corrections to the kinetic energy become important and the non-relativistic description breaks down. The {\bf gray dashed} lines show the NLO contribution to the UV phase in the Born approximation as defined in Eq.~\eqref{eq:NLOphaseshift}, normalized with respect to the leading order one. In the {\bf gray} shaded area the NLO contribution becomes larger than the LO one and the calculability of the phase breaks down. {\bf Left:} The {\bf blue} contours show the behavior of the mean value of the NLO SE normalized with respect to the leading order SE. {\bf Right:} The {\bf blue} contours show the behavior of the theory error on the NLO SE due to the uncertainty on the UV phase determination as defined in Eq.~\eqref{eq:SEerrorphase}.}
\label{fig:UV_plots}
\end{figure*}

The existence of a solution for Eq.~\eqref{eq:matching} in the first branch is intimately related to the calculability of the UV phase. For a solution to exist, the l.h.s of the equation must exceed the minimum of the r.h.s which is obtained for $\Phi_v\rightarrow 0$. This is only possible if $\bar{\delta}_0^{\text{NLO}}>0$, meaning that the leading-order (LO) Coulomb potential, which contributes positively to the scattering phase, must dominate over the NLO corrections, which contribute negatively (as seen in Eq.~\eqref{eq:prateek}). In the gray region in Fig.~\ref{fig:UV_plots}, finding a solution to the matching equation requires increasing $x_{\text{cut}}>1$, hence enhancing the relative weight of the Coulomb potential with respect to the NLO corrections. This approach allows us to extract the theoretical prediction for the SE and its associated uncertainties in this regime, as shown in Fig.~\ref{fig:abund_all}.    

Before leaving this section we comment on the UV NLO contributions to the BSF cross-section. These corrections affect both the deformation of the initial scattering wave function and the bound state wave function. While computing the latter is straightforward, in order to compute the first we generalize to discontinuous potentials the variable phase method introduced in \cite{Ershov:2011zz} (see also \cite{Beneke:2014gja,Asadi:2016ybp,Mahbubani:2020knq}). This is described in Appendix  \ref{app:VPM}. 

In general, we expect the NLO UV corrections to affect less BSF than the SE. In fact, while the SE is sensitive to the dynamics at the origin, which is dominated by the UV nature of the potentials, the relevant scale for BSF is given by the bound state Bohr radius, $a_B=2/\alpha_\chi M_\chi$. At this scale, the UV corrections are typically a sub-leading to the potential. For completeness, in Appendix \ref{app:BSF_UV} we show the effect of the NLO UV corrections to BSF in the model of Sec.~\ref{sec:darkQED} .

\section{An abelian example}\label{sec:darkQED}

We consider a dark $U(1)$ gauge theory where a pair of heavy scalar DM particles $\chi$ with mass $M_\chi$ and charge $q_\chi$ annihilates into a pair of dark photons. The leading order cross-section is $\sigma_h^0\vrel=2\pi \alpha^2_\chi/M_\chi^2$, where  $\alpha_\chi=g^2q_\chi^2/4\pi$ is the gauge coupling of the dark abelian gauge group evaluated at the DM mass. We also allow the presence of a light fermion $e$ of mass $m_e\ll M_{\chi}$ and charge $q_e$. The DM annihilates in dark sector states whose dynamics after DM freeze-out we ignore. In general, one should provide a mechanism for a quick and harmless decay of these states (see Ref.~\cite{Pospelov:2007mp,Feng:2008mu} for dedicated studies on secluded DM scenarios).

The scalar DM mainly annihilates into a pair of dark photons in $s$-wave, while the annihilation into light fermions is velocity suppressed. Hence, at leading order, the hard cross-section is dominated by the $j=0$ partial wave.\footnote{Focusing on scalar DM  simplifies both the spin structure of the hard cross-section and the one of the NLO potential, setting to zero all the spin-dependent contributions. For fermionic DM the computation of the SE is technically more complicated but the main message of this paper on how to assign a reliable theory uncertainty to the freeze-out masses is unchanged.} At leading order, the SE is just $S_E \simeq2\pi \alpha_\chi/\vrel$ for $v\ll\alpha$. This leads to a coefficient $c_\text{ann}^0=2$ as defined in Eq.~\eqref{eq:full}. BSF is dominated by the BS with principal quantum numbers $n\leq2$, since for large $n$ the BSF cross-section is suppressed by $\sim n^{-5/2}$ (see Appendix~\ref{sec:BSFNLO}). The dominant BSF channel is the formation of $1s$ and $2s$ bound states from a p-wave scattering state and $2p$ bound state from both s-wave and d-wave scattering state.\footnote{In the notation of Eq.~\eqref{eq:full} we can write the BSF contributions as $c_\text{BS}^1=512(e^{-4}+8e^{-8})/3\simeq 3.6$, $c_{\rm BS}^0=1024 e^{-8}/9\simeq0.04$ and $c_{\rm BS}^2=32768 e^{-8}/9\simeq1.2$.} The perturbative unitarity bound is dominated by the $j=0$ wave. From Eq.~\eqref{eq:PUBLO} we find $\alpha_{\rm LO}^{\max}=0.69$. By  using Eq.~\eqref{eq:mmax}, this implies $M_{\rm LO}^{\max}=240$~TeV as the perturbative unitarity bound on the scalar DM mass at leading order. 
 
We now want to estimate the accuracy of the theoretical prediction on the DM freeze-out mass approaching the perturbative unitarity bound. As the coupling strength increases, we expect the theory uncertainty to grow.  The leading corrections are likely to affect the Sommerfeld Enhancement (SE) and Bound State Formation (BSF), as these make the largest contributions to the annihilation cross-section in the non-relativistic limit. The origin of the theory uncertainty is not immediately apparent from the leading order computation. This is because the leading order Coulomb potential allows solutions that are regular everywhere and hence insensitive to the UV behavior of the theory. 

However, as discussed in Sec.~\ref{sec:NLO}, introducing NLO threshold corrections induced by heavy DM loops the SE at NLO becomes UV sensitive to the boundary conditions set by the UV scattering phase. The calculability of the scattering phase is challenged by the UV Landau pole for the dark $U(1)$ and its uncertainty dominates the theory error on the SE. 

For this simple theory, the NLO potential reads
\begin{equation}
    V_{\text{NLO}}(r)=V_{\text{NLO}}^{\rm IR}(r)+\Delta V_{\text{NLO}}^{\rm UV}(r)\ ,\label{eq:fullNLO}
\end{equation}
where the NLO contributions are \cite{Beneke:2019qaa,Beneke:2020vff,Urban:2021cdu,Beneke:2013jia}
\begin{align}
&V^{\text{NLO}}_{\rm IR}(r)=-\frac{\alpha(r)}{r},\quad \text{with}\quad \\
&\alpha(r)=\frac{\alpha_\chi}{1+\frac{\alpha}{2\pi} b^{\rm IR}\left[\log(m_e r)\theta(1-m_e r)+\log\left(\frac{M_\chi}{m_e}\right)\right]},\notag
\end{align}
and 
\begin{equation}\label{eq:pot}
\Delta V^{\text{NLO}}_{\rm UV}(r)=\frac{\alpha^2_\chi}{4M_\chi r^2}-\frac{7\alpha^2_\chi}{6\pi M_\chi^2}\text{reg}\frac{1}{r^3}\ .
\end{equation} 
The contribution of the light fermion to the $U(1)$ running is encoded in $b^{\text{IR}}=4q_e^2/3$. 

The main result of our analysis are shown in Fig.~\ref{fig:UV_plots}. The grey contours display the phase shift of the UV scattering phase due to the NLO correction, $\Delta\bar{\delta}_0^{\text{NLO}}$ which we define as 
\begin{equation}
\bar{\delta}_0^{\text{NLO} \pm}\equiv\bar{\delta}_0^{\text{LO}}+\Delta\bar{\delta}_0^{\text{NLO}}\pm (\Delta\bar{\delta}_0^{\text{NLO}})^2/\bar{\delta}_{\text{LO}}\ ,\label{eq:NLOphaseshift}
\end{equation}
where $\bar{\delta}_0^{\text{LO}}$ is the UV scattering phase due to the leading order Coulomb potential and $\bar{\delta}_0^{\text{NLO}}$ the mean value of the NLO one. Both phases are computed in the Born approximation plugging in Eq.~\eqref{eq:prateek} the full NLO potential in Eq.~\eqref{eq:fullNLO}. The only difference is that the short distance integral is truncated at $r_*\vert_{\text{max}}$ as defined in Sec.~\ref{sub:IR} to avoid the UV Landau pole of the dark $U(1)$. From Eq.~\eqref{eq:NLOphaseshift}, we defined the theory uncertainty interval in the determination of the UV scattering phase as $\bar{\delta}_0^{\text{NLO} +}-\bar{\delta}_0^{\text{NLO} -}=2(\Delta\bar{\delta}_0^{\text{NLO}})^2/\bar{\delta}_{\text{LO}}$.

As long as $\Delta\bar{\delta}_0^{\text{NLO}}/\bar{\delta}_0^{\text{LO}}<1$, the NLO prediction on the SE corresponds to the central value of the NLO UV phase in Eq.~\eqref{eq:NLOphaseshift}. This is shown by the blue contours in Fig.~\ref{fig:UV_plots} left. The theory error on the scattering phase induces an uncertainty in the determination of the SE
\begin{equation}
\Delta S_E^{\text{NLO}}\equiv S_E\vert_{\bar{\delta}_0^{\text{NLO} +}}-S_E\vert_{\bar{\delta}_0^{\text{NLO} -}}\, ,\label{eq:SEerrorphase}
\end{equation}
which is shown by the blue contours in Fig.~\ref{fig:UV_plots} right. Of course, when $\Delta\bar{\delta}_0^{\text{NLO}}/\bar{\delta}_0^{\text{LO}}>1$, the phase becomes incalculable as well as the associated NLO SE. This region is shaded in gray in Fig.~\ref{fig:UV_plots} and its onset signals the breakdown of calculability which starts well before the LO PUB at $\alpha_{\rm LO}^{\max}=0.69$ is achieved.

\begin{figure*}[htp!]
\centering
\includegraphics[scale=.3]{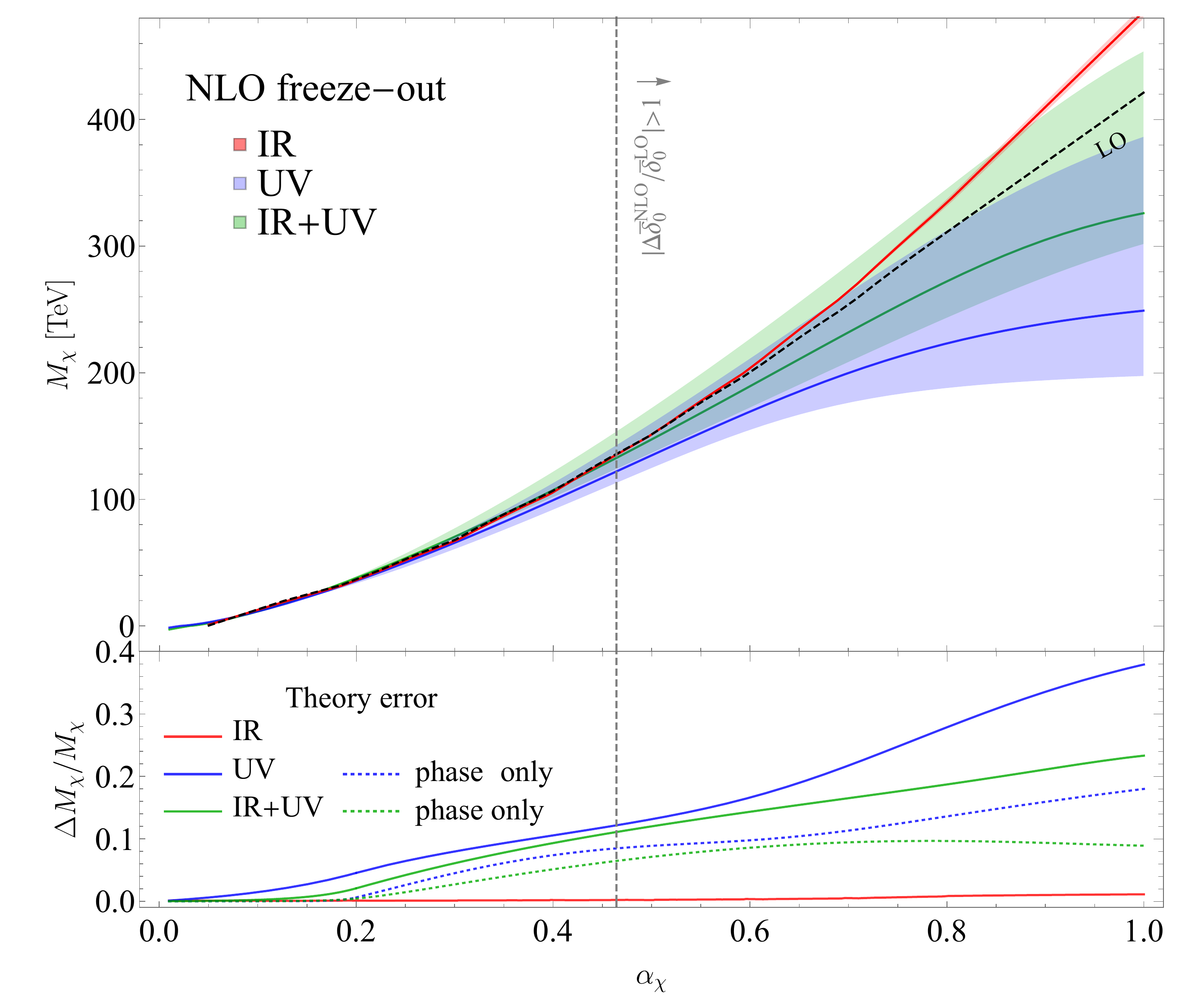}\qquad
\includegraphics[scale=.28]{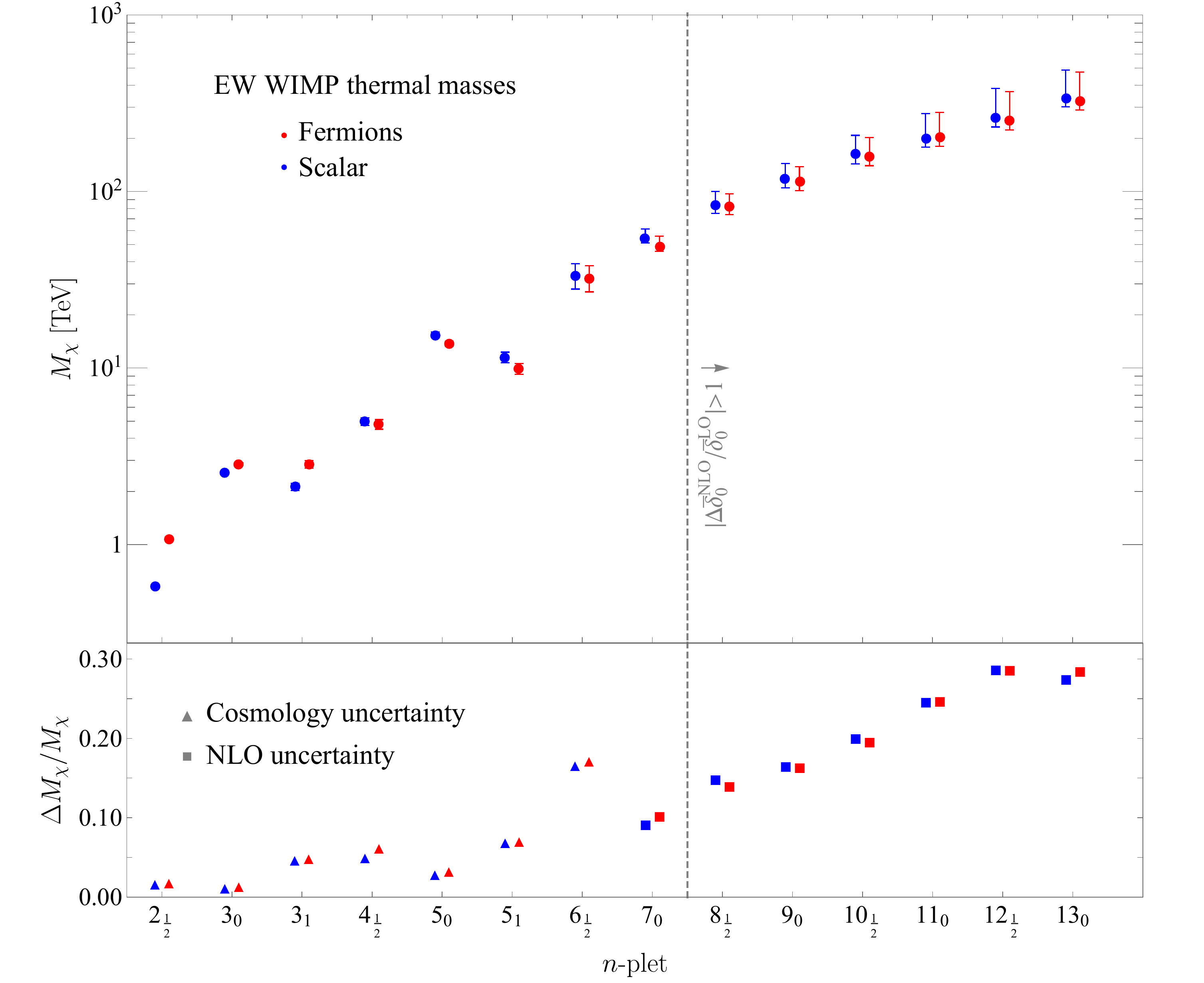}
\caption{{\bf Left:} DM thermal masses for leading order and NLO corrections to the Coulomb potential in the dark QED model introduced in Sec.~\ref{sec:darkQED} with $q_\chi=q_e=1$. The {\bf black} line shows the leading order prediction of the relic abundance. The {\bf red}, {\bf blue}, and {\bf green} lines show the freeze-out prediction after the inclusion of IR, UV, and IR+UV corrections respectively with their theoretical uncertainty shown as shaded regions of the same color. The {\bf gray dashed} line indicates the value of the coupling above which the UV phase becomes incalculable with $x_{\text{cut}}=1$. In the {\bf bottom} panel we show the theoretical uncertainty on the freeze-out prediction as a function of the coupling $\alpha$. The uncertainty is dominated by the UV NLO corrections ({\bf blue}) while the IR corrections ({\bf red}) are negligible in a IR-free theory. The IR+UV error {\bf green} is smaller because of a partial cancellation in the UV phase between the two contributions. We separately show the error due to the uncertainty in the UV phase defined in Eq.~\eqref{eq:SEerrorphase} ({\bf dotted} lines) and the total error which includes the systematic uncertainty of our regularization procedure with a single potential well-estimated in App.~\ref{sec:coulomb} ({\bf solid} lines). {\bf Right:} Estimated theory uncertainties on the EW $n$-plets thermal masses as computed in Ref.~\cite{Bottaro:2021snn,Bottaro:2022one}. The bottom panel shows that the relative uncertainty on the freeze-out mass is dominated by simplifications in the treatment of BS cosmology for $n<7$, while for $n\geq7$ the theory uncertainty is dominated by the NLO corrections to the annihilation cross-section discussed here.}
\label{fig:abund_all}
\end{figure*}

In Fig.~\ref{fig:abund_all} we show the leading order and NLO freeze-out predictions for this simple model setting $q_\chi=q_e=1$. As expected, both NLO corrections modify mostly the short-distance behavior of the Coulomb potential. Specifically, the IR NLO corrections tend to increase the SE by making the effective coupling larger at short distances, while the UV NLO corrections reduce the strength of the Coulomb potential hence reducing the SE. Combining the two effects make the mean of the NLO prediction accidentally close to the leading order one. 

The theory error on the freeze-out mass reflects the uncertainty in the determination of the UV phase.  A further intrinsic error arises from approximating the short distance potential as a single well matching solely the s-wave UV scattering phase $\bar{\delta}_0$. In principle, the procedure of Sec.~\ref{sec:UV} can be easily generalized by introducing an arbitrary number of potential wells at short distances with depths fixed by matching the UV theory scattering phases in different $l$-waves. We give an example of this generalized procedure for the leading order Coulomb potential in App.~\ref{sec:coulomb}. In the Coulomb case, we find that the SE computed using a single potential well, $S_E^{1w}$, underestimates the full leading order SE by a factor $\epsilon_{1w}\equiv S_E^{1w}/S_E^{\text{full}}$ which is explicitly shown in the blue line of Fig.~\ref{fig:three_counterterms}. Given that we are perturbatively expanding around the Coulomb case, we account for this extra ``systematic'' uncertainty by rescaling the upper limit on the theory error on the NLO SE defined in Eq.~\eqref{eq:SEerrorphase}: $S_E\vert_{\bar{\delta}_0^{\text{NLO} +}}\to S_E\vert_{\bar{\delta}_0^{\text{NLO} +}}/\epsilon_{1w}$. This procedure should conservatively account for all the theory uncertainties in our freeze-out prediction.

\subsection{Theory uncertainty on the electroweak WIMPs}

As an application of the previous computation, we go back to the freeze-out predictions for the electroweak WIMPs derived in Ref.~\cite{Bottaro:2021snn,Bottaro:2022one}. In principle, the computation of the previous section should be repeated for the $SU(2)$ $n$-plets, with the book-keeping challenge of including the NLO corrections in all the isospin channels for the SE and the BSF. Fortunately,  enlarging the representation $n$ of the multiplet, the non-relativistic potential is dominated by the abelian part. Therefore, we can approximately estimate the theory uncertainty for the EW WIMPs by taking the curve in the bottom plot of Fig.~\ref{fig:abund_all} left and rescaling the coupling constant $\alpha_\chi\to \alpha_{\text{eff}}\simeq (2n^2-1) \alpha_2/8$ focusing on the zero isospin channel. We plot this in Fig.~\ref{fig:abund_all} right, where we see that for EW multiplets with $n>7$ the dominant theory uncertainty comes from the NLO corrections while for lighter multiplets the theory error is dominated by the approximations on the bound state cosmology.\footnote{The error on small multiplets is dominated by our approximate treatment of BSF in the presence of EW interactions. In particular, the uncertainty comes from neglecting the masses of the EW gauge bosons and the details of bound state decoupling (see Ref.~\cite{Bottaro:2021snn}). For $n=6$ and for all $n>7$ we further neglected the bound state ionization from the thermal plasma. This introduces an additional error in the DM mass which was estimated to be at most $5$ TeV in Ref.~\cite{Bottaro:2021snn}. For $n=6$, this turns out to be the dominant source of error explaining the offset of the theory error for this multiplet.} Of course, this rough rescaling is far from being a full NLO computation for the EW WIMPs, but it already fixes the unphysical behavior of the theory uncertainty for large $n$-plets estimated in Ref.~\cite{Bottaro:2021snn,Bottaro:2022one}.

\section{Conclusions}\label{sec:conclusions}
In this paper, we studied NLO corrections to the freeze-out annihilation in the non-relativistic limit. These are important for heavy DM candidates, where the coupling strength approaches the perturbative unitarity bound as originally defined in Ref.~\cite{Griest:1989wd}. We discussed mostly IR-free gauge theories where the inclusion of NLO corrections makes it apparent that approaching the perturbative unitarity bound, the theory uncertainty due to the matching of the UV data onto the non-relativistic NLO potential blows up. This allows us to estimate a trustworthy theory error on the heavy EW WIMPs masses computed in Ref.~\cite{Bottaro:2021snn,Bottaro:2022one}. %

We defined a systematic procedure to perform the matching from the UV relativistic theory to the non-relativistic potential. Even though our work builds upon previous works on the subject we believe that many subtle issues were clarified here. We hope that this can serve as a basis for further studies in this direction. For instance, we leave for the future a systematic treatment of NLO corrections in UV free gauge theories and a careful assessment of the impact of NLO corrections in exclusive channels like the ones considered in indirect detection~\cite{Baumgart:2017nsr, Cirelli:2015bda, Garcia-Cely:2015dda}.

\begin{acknowledgments}
We thank Roberto Franceschini for asking how to reliably assign a theory error to freeze-out predictions. We are grateful to Brando Bellazzini for many enlightening discussions about Ref.~\cite{Bellazzini:2013foa}. We also thank Prateek Agrawal and Aditya Parikh for discussions about Ref.~\cite{Agrawal:2020lea}. We thank Neot Smadar, CERN, and the Galileo Galilei Institute for hospitality during the completion of this work. We thank Marco Costa and Nick Rodd for a careful read and useful feedback on the draft. SB is supported by the Israel Academy of Sciences and Humanities \& Council for Higher Education Excellence Fellowship Program for International Postdoctoral Researchers.
\end{acknowledgments}

\bibliographystyle{JHEP}
\bibliography{SE_NLO}

\newpage
\appendix
\onecolumngrid

\section{The UV potential}\label{app:fourier}

The UV potential in Eq.~\eqref{eq:NLO_UV} has been computed for any SU(N) theory up to three-loop order in \cite{Beneke:2013jia}. The proper setup for the computation is the so-called potential non-relativistic QFT (pNRQFT). This is obtained from the standard NRQFT by integrating out vector bosons with energy of order $M_\chi \vrel$. In this EFT, which is local in time but non-local in space, the potentials emerge as the coefficients of the 4-point functions of the heavy non-relativistic fields (the DM field in our case). In Fourier space, these potentials can be organized in powers of $1/M_\chi$ as follows:

\begin{equation}\label{eq:pot_general}
\begin{split}
\V_{\rm NLO}^{\rm UV}(p)=&-\V_C(\alpha)\frac{4\pi q_\chi^2 }{p^2}+\V_{1/M_\chi}(\alpha)\frac{\pi^2(4\pi) q_\chi^2\alpha }{\mdm p}+\V_\delta(\alpha)\frac{2\pi q_\chi^2  \alpha}{\mdm^2}+\V_S(\alpha)\frac{4\pi q_\chi^2 \alpha}{3\mdm^2}\vec{S}^2\\&+\V_{\text{hf}}(\alpha)\frac{\pi q_\chi^2 \alpha}{3M_\chi^2p^2}(p^2\delta_{ij}-3p_ip_j)\sigma_{1i}\sigma_{2j}-\V_{\text{so}}(\alpha)\frac{6i\pi q_\chi^2 \alpha}{M_\chi^2p^2}(\vec{p}\times\Vec{q})\cdot \vec{S},
\end{split}
\end{equation}

where $\vec{s}_i=\vec{\sigma}_i/2$ is the spin of particle $i$ and $\vec{S}=\vec{s}_1+\vec{s}_2$ is the total spin of the scattering state. We neglected additional, velocity-suppressed, $\mathcal{O}(1/M_\chi^2)$ terms. Each $\V_i(\alpha)$ can be expanded in powers of $\alpha$ $\V_i(\alpha)=\sum_{L=0}^\infty\left(\alpha/4\pi\right)^L\V_i^{(L)}$
where $L$ is the loop order. The Coulombian term $\V_C(\alpha)$ simply reproduces the running coupling below the $M_\chi$ threshold and needs no further discussion. Here we report the rest of the $\V_i(\alpha)$ coefficients evaluated at one-loop \cite{Beneke:2013jia}:

\begin{equation}
\begin{split}
&\V_{1/M_\chi}(\alpha)=\frac{q_\chi^2\alpha}{8\pi}\ ,\\
&\V_{\delta}(\alpha)=\frac{13q_\chi^2\alpha}{15\pi}-\frac{4q_\chi^2\alpha}{3\pi}\log\left(\frac{\mu^2}{M_\chi^2}\right)-\frac{7q_\chi^2\alpha}{6\pi}\log\left(\frac{M_\chi^2}{p^2}\right)\ ,\\
&\V_s(\alpha)=1-\frac{\alpha}{\pi}\left(\frac{1}{2} q_\chi^2+\frac{5}{9}q_e^2+\frac{1}{3}q_e^2\log\frac{\mu^2}{p^2}\right)\ ,\\
&\V_{hf}(\alpha)=1+\frac{\alpha}{\pi}\left(q_\chi^2-\frac{5}{9}q_e^2-\frac{1}{3}q_e^2\log\frac{\mu^2}{p^2}\right)\ ,\\
&\V_{so}(\alpha)=1+\frac{\alpha}{3\pi}\left(1-\log\frac{\mu^2}{p^2}\right)\ .
\end{split}
\end{equation}

For scalar DM $(s=0)$ in an s-wave scattering state $(l=0)$ only the first row in \eqref{eq:pot_general} contribute, so that we get 
\begin{equation}
\V_{l=0}=-\frac{4\pi q_\chi^2\alpha}{p^2}+\frac{\pi^2q_\chi^2\alpha^2}{2M_\chi p}+\frac{2\pi q_\chi^2\alpha}{M_\chi^2}\left(\frac{13q_\chi^2\alpha}{15\pi}-\frac{4q_\chi^2\alpha}{3\pi}\log\left(\frac{\mu^2}{M_\chi^2}\right)-\frac{7q_\chi^2\alpha}{6\pi}\log\left(\frac{M_\chi^2}{p^2}\right)\right)
\end{equation}

In order to convert the potentials above into real space we need the Fourier transforms
\begin{equation}
F\left(\frac{1}{p}\right)=\frac{1}{2\pi^2r^2}\quad ,\quad F\left(\log(p^2)\right)=-\frac{1}{2\pi}\text{reg}\frac{1}{r^3}\ ,
\end{equation}
where the regularized $1/x^n$ distribution is defined as
\begin{equation}
\int \mathrm{d}^n x \phi(\vec{x})\text{reg}\frac{1}{x^n}=\lim_{\epsilon\rightarrow 0}\left[\int \mathrm{d}^n x \phi(\vec{x})\frac{x^\epsilon}{x^n}-\phi(0)A(n,\epsilon)\right]
\end{equation}
with
\begin{equation}
\begin{split}
&A(n,\epsilon)=\frac{2\pi^\frac{3}{2}}{\Gamma\left(\frac{n}{2}\right)}\left(\frac{1}{\epsilon}+\log 2+\frac{\psi\left(\frac{n}{2}\right)-\gamma_E}{2}\right),\\
&\psi(x)=\frac{\mathrm{d}\log\Gamma(x)}{\mathrm{d}x}\ .
\end{split}
\end{equation}
After Fourier transform we get
\begin{equation}
V_{\rm NLO}^{l=0}(r)=-\frac{ q_\chi^2\alpha(r)}{r}+\frac{q_\chi^4\alpha^2}{4M_\chi r^2}-\frac{7q_\chi^4\alpha^2}{6\pi M_\chi^2}\text{reg}\frac{1}{r^3}+\frac{2 q_\chi^4\alpha^2}{3M_\chi^2}\left(\frac{13}{5}-4\log\left(\frac{\mu^2}{M_\chi^2}\right)\right)\delta^{(3)}(\vec{r})
\end{equation}

In the general case, the potential \eqref{eq:pot_general} preserves only the total angular momentum $\vec{J}=\vec{L}+\vec{S}$. In particular, the hyperfine $\V_{\text{hf}}$ and the spin-orbit $\V_{\text{so}}$ terms will couple states with different $l$ and $s$ but with the same total angular momentum $j$. Extracting the SE in this case would require a straightforward generalization of the procedure outlined in Section \ref{sec:NLO}. 

\section{UV matching beyond single well}\label{sec:coulomb}

\begin{figure}[htp!]
    \centering
    \includegraphics[scale=0.5]{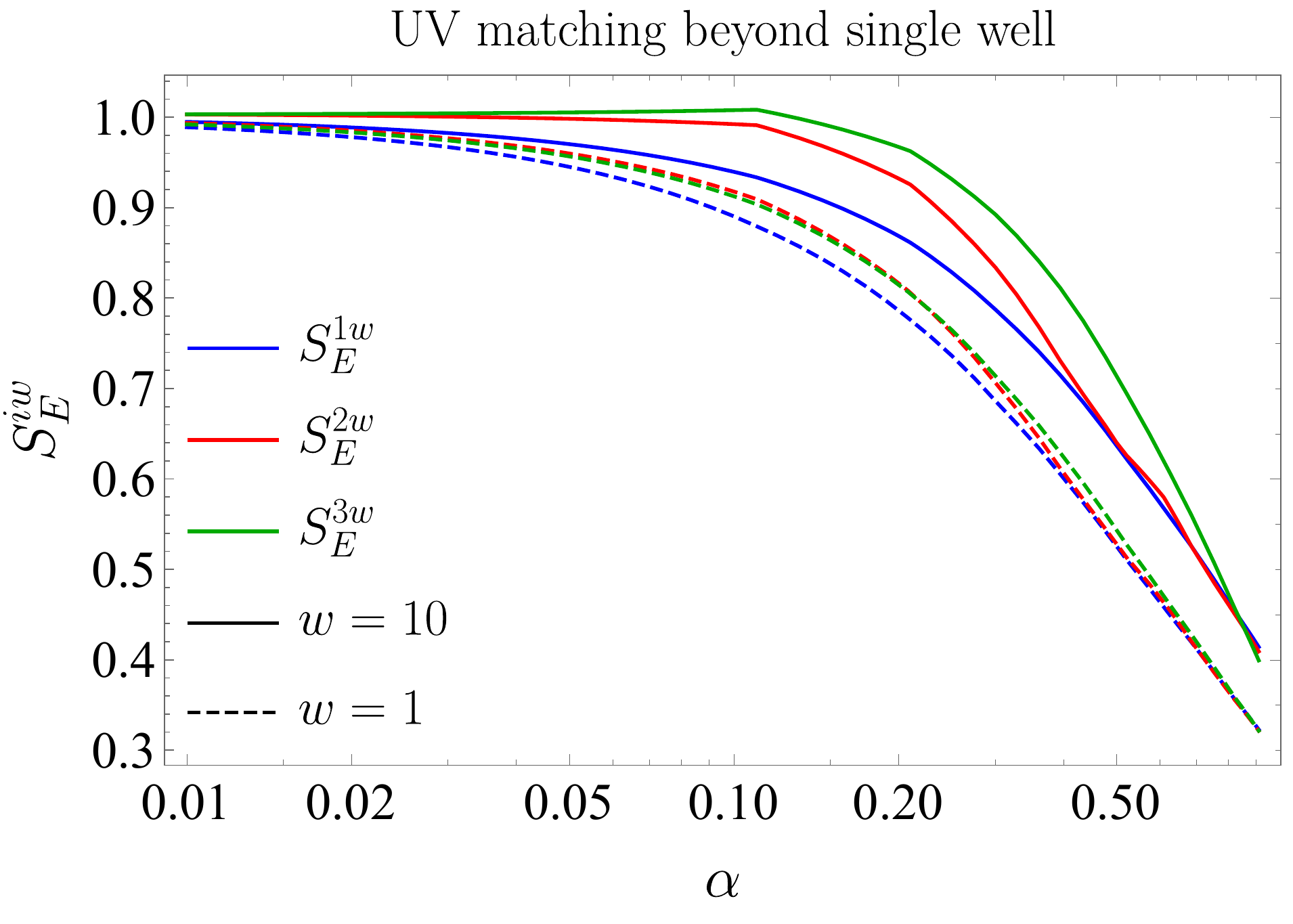}
    \caption{Ratio between the SE computed with our UV regularization procedure $S_E^{iw}$ with $i$ potential well and the full SE for the Coulomb potential as a function of the coupling strength $\alpha$ and fixing $v=0.02$. The {\bf blue} line corresponds to the single potential well with $x_{\text{cut}}=1$, the {\bf red} line corresponds to two potential wells with $x_1=1/6$ as defined in Eq~\eqref{eq:multiwell}. The {\bf green} line shows the result with three potential wells with $x_1=1/3$ and $x_2=2/3$. Solid (dashed) lines are obtained by choosing $w=10$ ($w=1$) in the determination of the wave function renormalization $Z_\chi$ as in \eqref{eq:zchi}.}
    \label{fig:three_counterterms}
\end{figure}

In this Appendix, we apply the procedure described in Sec. \ref{sec:UV} to the LO Coulomb potential. In Figure \ref{fig:three_counterterms} we show the ratio between the SE computed with the UV regularization procedure and the exact, analytical result in Eq.~\eqref{eq:som_en}. As we can see, regularizing the potential close to the origin leads to an underestimate of the SE by more than 50\% at couplings close to the LO PUB. 

One way to improve the precision of the calculation is to introduce more parameters to model the UV behavior of the potential. A very simple possibility is to use a series of well potentials:
\begin{equation}
    V_{\rm reg}^{(n)}(x)=-\sum_i V_i \varphi(x_{i-1},x_{i})\ ,\label{eq:multiwell}
\end{equation}
where we defined the function $\varphi(x_{i-1},x_{i})=1$ if $x_{i-1}<x<x_{i}$ and 0 otherwise, with $x_0=0$ and $x_n=x_{\rm cut}$. Just like the single well regularization, we need to find $n$ matching conditions such to fix the potential well depths $V_i$. These conditions are provided by the UV scattering phases for generic angular momenta up to $l=n-1$
\begin{equation}
\sin\bar{\delta}_l=-\frac{M_\chi^2 \vrel}{2}\int_0^{1/\mdm}\mathrm{d}r r^2 j_l(kr)^2 V(r),\quad k=\frac{M_\chi\vrel}{2}\ .
\end{equation}
Inside each well, the solution of the Schr\"{o}dinger equation
\begin{equation}
-\chi_i''(x)+\left(\frac{l(l+1)}{x^2}-\frac{4V_i+v^2}{4 Z_\chi}\right)\chi_i(x)=0
\end{equation}
is given by the combination:
\begin{equation}
\label{eq:sol_wells}
    \chi_i(x)=c_{j_li}\tilde{j}_l(\Phi_i x)+c_{y_li}\tilde{y}_l(\Phi_i x)\  ,\quad \Phi_i^2\equiv\frac{4V_i+v^2}{4 Z_\chi}\ ,
\end{equation}
 where $\tilde{j}_l(x)\equiv x j_l(x)$ $\tilde{y}_l(x)\equiv x y_l(x)$ with $j_l(x)$ and $y_l(x)$ being the regular and irregular spherical Bessel functions. The coefficients $c_{j_li}$ and $c_{y_li}$ are fixed by matching the solutions Eq.~\eqref{eq:sol_wells} at the boundaries of each potential well, where the coefficients $c_{j_l1}=1$ and $c_{y_l1}=0$ are determined by the boundary condition $\chi(x\rightarrow 0)\propto x^{l+1}$. The matching condition to the $l$-th scattering phase reads
\begin{equation}
\frac{\tilde{j}_l\left(\frac{v\xcut}{2}+\bar{\delta}_l\right)}{\tilde{j}_l'\left(\frac{v\xcut}{2}+\bar{\delta}_l\right)}=\frac{v}{2\Phi_n}\frac{c_{j_ln}\tilde{j}_l\left(\xcut\Phi_n\right)+c_{y_ln}\tilde{y}_l\left(\xcut\Phi_n\right)}{c_{j_ln}\tilde{j}_l'\left(\xcut\Phi_n\right)+c_{y_ln}\tilde{y}_l'\left(\xcut\Phi_n\right)}\ ,
\end{equation}
where both $c_{j_ln}$ and $c_{y_ln}$ are functions of all the $\Phi_i$'s. Solving the system above gives $\Phi_i(v,Z_\chi)$. The wave function renormalization $Z_\chi$ is fixed like in the single well potential by requiring that the s-wave SE with the regularized potential is 1 at large velocities, in the sense discussed in Section \ref{sec:UV}.\\
We show the results in Fig. \ref{fig:three_counterterms}. Blue, red and green lines represent the ratio between the SE computed with the potential regularized using one, two and three wells, respectively, and the exact result. The solid lines correspond to the value of this ratio as a function of $\alpha$ choosing $w=10$ as the large velocity at which we extract the wave function renormalization  from the solution of \eqref{eq:zchi}. The dashed lines show the same result employing the more physical $w=1$. As we can see, the agreement is significantly better for $w=10$, as we argued in \ref{sec:UV}, and the convergence towards the exact result increasing the number of counterterms is much faster as compared to $w=1$. 

\begin{figure}[htp!]
    \centering
    \includegraphics[scale=0.5]{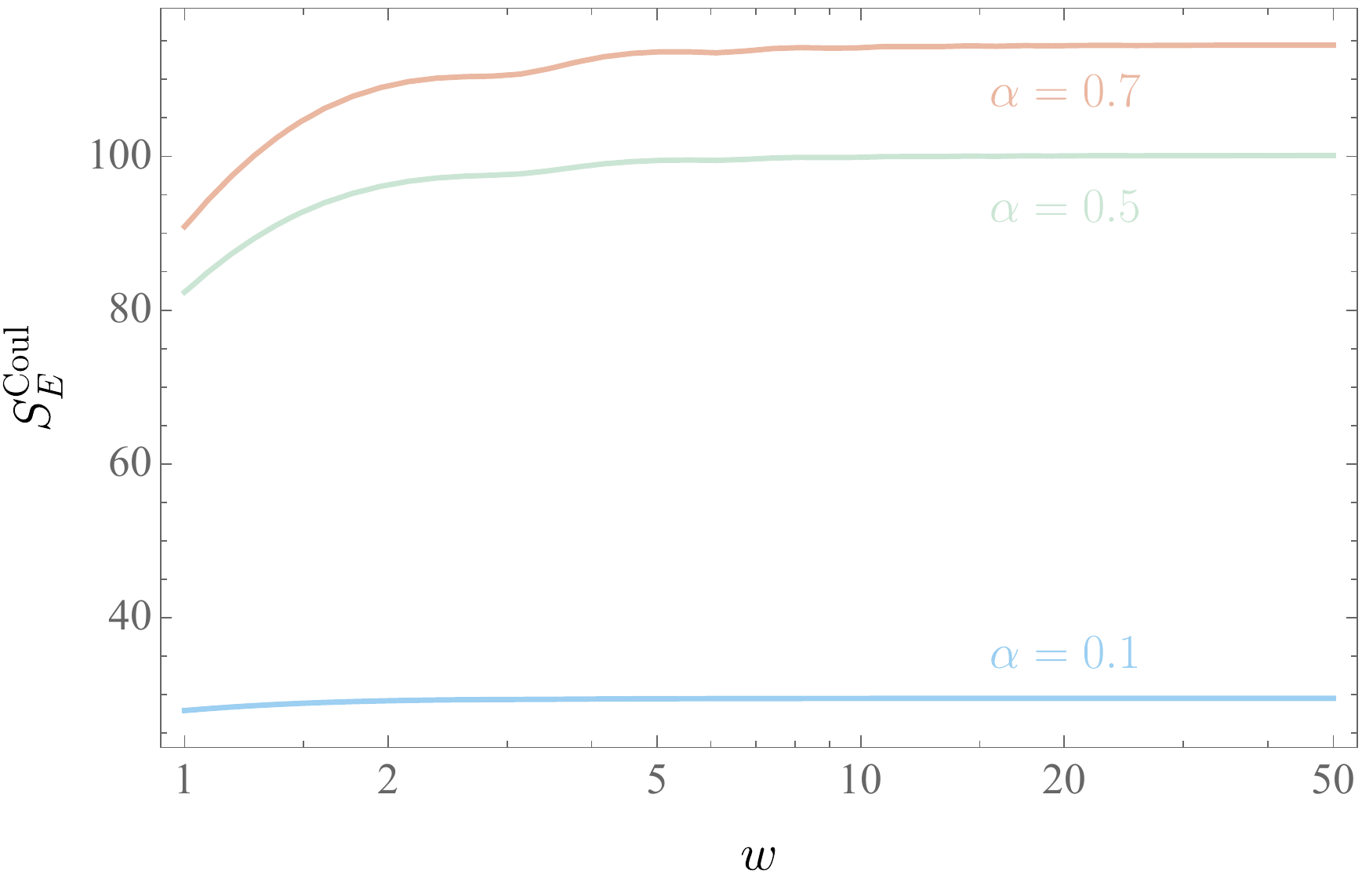}
    \caption{Residual spurious dependence of the extracted Sommerfeld enhancement on the velocity $w$ used to impose the renormalization condition as in Eq.~\eqref{eq:boundarygeneral}. }
    \label{fig:referee2}
\end{figure}

Another way to motivate the boundary condition in Eq.~\eqref{eq:bc2} is to generalize it for an arbitrary velocity by imposing
\begin{equation}
S_E(w)=1\ ,\label{eq:boundarygeneral}
\end{equation}
and then study how much the Sommerfeld enhancement preserves a spurious dependence on the precise value of the velocity $w$ at which the wave function renormalization is extracted. In Fig~\ref{fig:referee2} we show such a residual dependence, which as expected disappears only for large enough $w$.

\section{More about Bound State Formation}\label{sec:BSFNLO}
In this section, we summarize useful formulas for BSF focusing on abelian gauge theories. In Sec.~\ref{sec:BSFLO} we first review the basics of BSF at LO (more details can be found in Ref.~\cite{vonHarling:2014kha,Mitridate:2017izz,Harz:2018csl,Bottaro:2021snn}). In Sec.~\ref{sec:scaling} we give a simple argument for abelian gauge theories showing the decoupling of BSF at large BS quantum numbers. In Sec.~\ref{sec:NLOBSF} we consider NLO corrections to the BSF showing that are subleading with respect to the NLO corrections to the SE considered here. Similar results were presented in Ref.~\cite{Bottaro:2021snn}. In Sec.~\ref{app:VPM} we present a generalization of the variable phase method that we used to compute the BSF cross-sections with NLO potentials. 
\subsection{LO Bound State Formation}\label{sec:BSFLO}
At leading order, the bound state formation proceeds in the electric dipole approximation through the emission of a single vector boson $\gamma$ as $\chi(\vec{p}_1)+\bar{\chi}(\vec{p}_2)\rightarrow \mathrm{BS}(\vec{q})+\gamma(\vec{k})$. This process is described by the effective Hamiltonian with the usual electric dipole term

\begin{equation}\label{eq:BS_darkQED}
    \mathcal{H}^{\text{LO}}=-\frac{g q_\chi}{M_\chi}(\vec{\gamma}^a(x_1)\cdot \vec{p}_1-\vec{\gamma}^a(x_2)\cdot \vec{p}_2)\ .
\end{equation}

A BS with principal quantum number $n$, angular momentum $(l,m)$, spin $s$, labeled as $^s\mathcal{B}_{nl}$, is described by the wave function $\chibs(r)$ satisfying the Schr\"{o}dinger equation
\begin{equation}
\label{eq:Schrodinger_BS}
-\frac{\nabla^2\chibs}{M_\chi}+V_{i}^{\text{LO}}\chibs = E_{nl}\chibs\, ,
\end{equation}
where $E_{nl}$ is the binding energy of  the BS which for a Coulomb potential is $E_{nl}=q_\chi^4\alpha^2 M_\chi/4n^2$. The cross-section for the formation of a BS can be written as 

\begin{equation}\label{eq:bsf_xsec}
   \sigma_\text{BS}v=\!\frac{2\alpha q_\chi^2}{\pi}\frac{k}{M_\chi^2}\frac{2s+1}{g_\chi^2}\sum_{m}\int\!\!\mathrm{d}\Omega_k\left(|\vec{\mathcal{A}}_{p,nlm}|^2-\frac{|\vec{\mathcal{A}}_{p,nlm}\cdot \vec{k}|^2}{k^2+m_\gamma^2}\right),
\end{equation}

where $g_\chi$ indicates the number of DM degrees of freedom, including those coming from internal quantum numbers and $k$ is the momentum of the emitted vector, whose mass is $m_\gamma$. In particular, for massless vectors $k= E_{nl}+ M_\chi \vrel^2/4$. The amplitude can be written as

\begin{equation}\label{eq:overlap}
\vec{\mathcal{A}}_{p,nlm}=\!\int\mathrm{d}^3r\phi_{nlm}^*(r)\vec{\nabla}\phi_{p}(r)\ ,
\end{equation}
where  we defined  the scattering wave function $\phi_p$ satisfying Eq.~\eqref{eq:Schrodinger} with $p=M_\chi\vrel/2$. By inspecting the overlap integrals, we can provide a simple parametric estimate of the BSF cross-section in the $\vrel<\alpha$ limit
\begin{equation}
    \sigma_{\text{BS}}\vrel\simeq  \frac{8\pi^2\alpha q_\chi^2}{g_\chi^2 M^3_\chi\vrel}E_{nl}\ .
\end{equation}
The contribution of BSF to the DM annihilation cross-section is the result of balancing three different processes in the dynamics of a BS in the plasma: i) BS annihilation into plasma states, ii) BS decay or excitation to other BSs, iii) BS ionization due to the thermal plasma.  

The ionization rate in the thermal plasma can be related to the BSF cross-section in Eq.~\eqref{eq:bsf_xsec} through detailed balance:
 \begin{equation}
    \Gamma_\text{break}=\frac{g_\chi^2}{g_\text{BS}}\frac{(M_\chi T)^{\frac{3}{2}}}{16\pi^{\frac{3}{2}}}e^{-\frac{E_{nl}}{T}}\langle\sigma_\text{BS}v\rangle\ ,
\end{equation}
and it depends exponentially on the hierarchy of thermal bath temperature $T$ and the BS binding energy $E_{nl}$. 
The decay/excitation rate is computed by replacing the scattering wave function in the BSF with the wave function of the excited BS. For example, in the simple dark QED model considered in Sec~\ref{sec:darkQED}, the decay rate as a function of $n$ can be written explicitly as: 
\begin{equation}
    \Gamma_\text{dec}^{\text{U(1)}}\simeq \frac{(q_\chi^2\alpha)^5M_\chi}{3n^5}\ .
\end{equation}
As shown in Ref~\cite{Mitridate:2017izz,Harz:2018csl} the Boltzmann equations for the DM and the BS abundances can be enormously simplified whenever the rate of at least one of the three fundamental processes controlling BS dynamics is larger than the Hubble expansion. In this approximation, the effect of BSF can be encoded in an effective annihilation cross-section given by
\begin{equation}\label{eq:annEFF}
    \langle\sigma v\rangle=\langle\sigma_\text{ann} v\rangle+\sum_{\text{BS}}R_\text{BS}\langle\sigma_\text{BS} v\rangle,
\end{equation}
where $R_\text{BS}$ is the annihilation branching ratio of the BS which for a single bound state takes a rather intuitive form $R_\text{BS}=\Gamma_\text{ann}/(\Gamma_\text{ann}+\Gamma_\text{break})$ and for a general excited state BS is
\begin{equation}
    R_\text{BS}=\frac{\sum_i\Gamma_\text{dec}(\text{BS}\rightarrow \text{BS}_i) R_{\text{BS}_i}}{\sum_i\Gamma_\text{dec}(\text{BS}\rightarrow \text{BS}_i)+\Gamma_\text{break}}\ ,
\end{equation}
assuming a negligible excitation rate. These branching ratios effectively reduce the impact of excited states on the DM abundance making the sum in Eq.~\eqref{eq:annEFF} dominated by the BS's with low principal quantum number $n$. In the model of Sec.~\ref{sec:darkQED}, a very good approximation is to take $R_{\rm BS}=1$, especially at large coupling, because of the increasing Boltzmann suppression in the ionization rate. The contribution of the more excited BS, for which  $R_{\rm BS}=1$ may be no longer valid, decouples very fast as we will show in Sec.~\ref{sec:scaling}). We checked that the error done by neglecting all states with $n\geq 3$ is smaller than the NLO theoretical uncertainty discussed in Sec.~\ref{sec:darkQED}. 

\subsection{BSF decoupling at large BS quantum number}\label{sec:scaling}

Here we show that for abelian interactions the BSF contribution is always dominated by the formation of lowest-lying states regardless of the interactions with the plasma. First of all, for a given $n$ BSF is dominated by the formation of the state with the largest angular momentum, that is $l=n-1$. This is due to cancellations in the overlap integrals due to the oscillatory behavior of the BS radial wave function for $l<n-1$, which is absent for maximal $l$. In this latter case, in fact, the bound state wave function goes like $\phi_{n,n-1}\propto x^n e^{-x/n}$, where $x=r/a_B$ is the radial coordinate in units of the Bohr radius $a_B=\frac{2}{\alpha q_\chi^2 M_\chi}$. In the large $n$ limit, we can rewrite the wave function above using the saddle point approximation
\begin{equation}
    \phi_{n,n-1}\simeq N(n)e^{-\frac{1}{2n}\left(\frac{x}{n}-n\right)^2}\ ,
\end{equation}
where $N(n)=n^{-11/4}$ to ensure the correct normalization of the wave function. By explicitly writing the gradient in spherical coordinates, the BSF amplitude in Eq.~\eqref{eq:bsf_xsec} goes like
\begin{equation}
    \mathcal{A}_{p,n(n-1)}\simeq\int\mathrm{d}r r^2 \phi_p(r)\left(\partial_r-\frac{n}{r}\right)\phi_{n,n-1}(r)\simeq \int\mathrm{d}r r \cos(pr)\left(\partial_r-\frac{n}{r}\right)\phi_{n,n-1}(r)\ ,
\end{equation}
where, apart from some irrelevant phase, we took the asymptotic limit of the scattering radial wave function $\phi_p(r)\sim\cos(pr)/r$. At large $n$ the integral is dominated by the non-derivative term so that we can write for the BSF cross-section
\begin{equation}
    \sigma_{\rm BS}\vrel\propto E_B|\mathcal{A}_{p,n(n-1)}|^2\propto \left|\int\mathrm{d}r\cos(pr)\phi_{n,n-1}(r)\right|^2\leq  \left|\int\mathrm{d}r\phi_{n,n-1}(r)\right|^2\ ,
\end{equation}
where $E_B\propto 1/n^2$ is the BS binding energy. Plugging in the saddle point approximation of $\phi_{n,n-1}(r)$ we get
\begin{equation}
    \sigma_{\rm BS}\vrel\propto n^{-\frac{5}{2}}\ .
\end{equation}

\subsection{Additional NLO contributions}\label{sec:NLOBSF}

\begin{figure*}[htp!]
    \centering
\includegraphics[ width= 0.99 \linewidth]{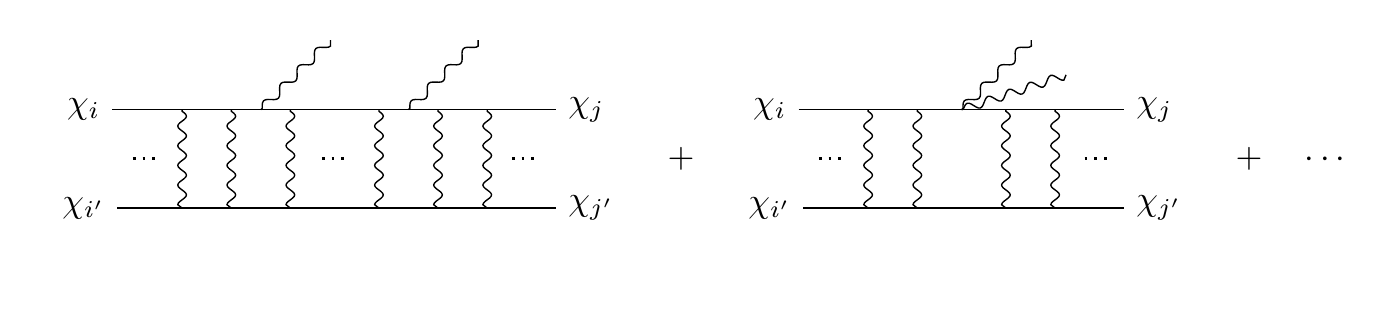}
    \caption{Examples of diagrams controlling the BS effective Hamiltonian at next-to-leading order in gauge boson emission The first diagram corresponds to the second order Born approximation for the dipole operators in \eqref{eq:BS_darkQED}, where the resummation of the vector boson insertions between the two emissions reconstructs the wave function of an intermediate BS or scattering state. The second diagram, instead, is obtained from the $\mathcal{O}(\gamma^2)$ terms in the interaction Hamiltonian, at leading order in the Born approximation.}
    \label{fig:abelian}
\end{figure*}

The main NLO contributions to BSF come from diagrams like the ones in Fig. \ref{fig:abelian} and are essentially of two types: i) the first diagram is essentially the second-order Born approximation of the LO Hamiltonian, with the intermediate state being a free or a BS; ii) the second diagram, where the two emitted vectors come from the same vertex, is generated by the effective Hamiltonian at order $\mathcal{O}(\gamma^2)$. The latter contains terms of the form
\begin{equation}
\Delta\mathcal{H}_I \supset \frac{g^2 q_\chi^2}{2M_\chi}\left[\vec{\gamma}\cdot\vec{\gamma} + \frac{(\vec{p}\cdot \vec{\gamma})(\vec{p}\cdot\vec{\gamma})}{M^2_\chi}\right],
\end{equation}
where we focus here on the abelian part of the Hamiltonian, postponing a full study for future work.  Given the above Hamiltonian, we can estimate the corresponding contribution to the double emission BSF cross-section as: 
\begin{equation}
    \Delta\sigma_{\text{BS}}\vrel\simeq \frac{\pi}{2g_\chi^2M_\chi^2\vrel}\left(\frac{E_{nl}}{M_\chi}\right)^3\ .\label{eq:NLObs}
\end{equation}

We now discuss the contributions from second-order Born expansion whose general expression is given by
\begin{equation}
\label{eq:sig2V}
(\sigma_{\rm BS} \vrel)_{2V} =\frac{2^6\alpha^2q_\chi^4}{3^3g_\chi^2\pi M_\chi^4} \int\!\!\mathrm{d}\omega \omega(E_n-\omega)\left|\mathcal{C}_{\text{BS}}+\mathcal{C}_{\text{free}}\right|^2\ ,
\end{equation}
where we defined 
\begin{subequations}
\label{eq:cint}
\begin{align}
&\mathcal{C}_{\text{BS}}=\sum_m\left(\frac{1}{E_n-E_m-\omega+i\Gamma_{\text{dec},m}}+\frac{1}{\omega-E_m+i\Gamma_{\text{dec},m}}\right)\mathcal{I}_{\vec{q}m}\mathcal{I}_{mn}\ ,\\
&\mathcal{C}_{\text{free}}=\int\frac{\mathrm{d}^3k}{(2\pi)^3}\left(\frac{1}{E_n-\omega+\frac{k^2}{M_\chi}+i\epsilon}+\frac{1}{\omega-\frac{q^2}{M_\chi}+\frac{k^2}{M_\chi}+i\epsilon}\right)\mathcal{I}_{\vec{q}\vec{k}}\mathcal{I}_{\vec{k}n}\ ,
\end{align}
\end{subequations}
with $\mathcal{I}_{if}$ being the overlap integrals between the states $i$ and $f$, the index $m$ running over all intermediate BS and the $k$-integral running over all the intermediate scattering states. 

Starting from $\mathcal{C}_{\text{BS}}$, the intermediate BS are rather narrow resonances because
\begin{equation}
\Gamma_{\text{dec}}\sim \alpha^3 E_{\rm BS} \ll E_{\rm BS}\ .
\end{equation}
 This quick estimate, supported by the full numerical computation, suggests that $\mathcal{C}_{\text{BS}}$ contribution is fully captured in the Narrow Width Approximation (NWA) for the intermediate BS. Therefore, neglecting the interference terms, one gets
\begin{equation}
(\sigma_{\rm BS} \vrel)_{2V}=\sum_m (\sigma_{\rm BS} \vrel)_{1V,m}\mathrm{BR}_{m\rightarrow n}\ ,
\end{equation}
which is exactly the single emission result.

To estimate the contribution from $\mathcal{C}_{\text{free}}$ we need to estimate $\mathcal{I}_{\vec{q}\vec{k}}$ which encodes the contribution from intermediate continuum states. For simplicity, we stick to the abelian contribution which reads 
\begin{equation}
\mathcal{I}_{\vec{q}\vec{k}}=\int r^2\mathrm{d}rR_{\vec{k}}\partial_r R_{\vec{q}}\ .
\end{equation}
The integral above can be split into small and large $r$ regions, roughly separated by the Bohr radius $a_B=\frac{1}{\alpha M_\chi}$
\begin{equation}
\begin{split}
\mathcal{I}_{\vec{q}\vec{k}}= &\int_0^{a_B} r^2\mathrm{d}rR_{\vec{k},1}\partial_r R_{\vec{q}}+\int_{a_B}^{\infty} r^2\mathrm{d}rR_{\vec{k}}\partial_r R_{\vec{q}}\\
\sim & \frac{1}{\alpha M_\chi \sqrt{k q}}+\frac{q}{(M_\chi \alpha)^2}\delta(q-k)\ ,
\end{split}
\end{equation}
which plugged into \eqref{eq:cint} gives an estimate to $\mathcal{C}_{\text{free}}$. All in all, plugging these estimates in \eqref{eq:sig2V} and replacing $q=M_\chi \vrel$ we get that the contribution from NLO exchange of continuum states behaves similarly to the ones estimated in \eqref{eq:NLObs}  up to subleading terms in the $\vrel<\alpha$ regime.

In conclusion, NLO corrections to BSF are suppressed by $\sim\alpha^3/64\pi$ with respect to the LO ones and can thus be neglected as compared to those considered in the main text. Other types of NLO corrections can be relevant at large temperatures, like BSF from scattering of fermions in the plasma, see \cite{Binder:2020efn} for a detailed analysis.

\subsection{Sensitivity of BSF to NLO potentials}\label{app:BSF_UV}

\begin{figure}[htp!]
\centering
\includegraphics[width=0.47\textwidth]{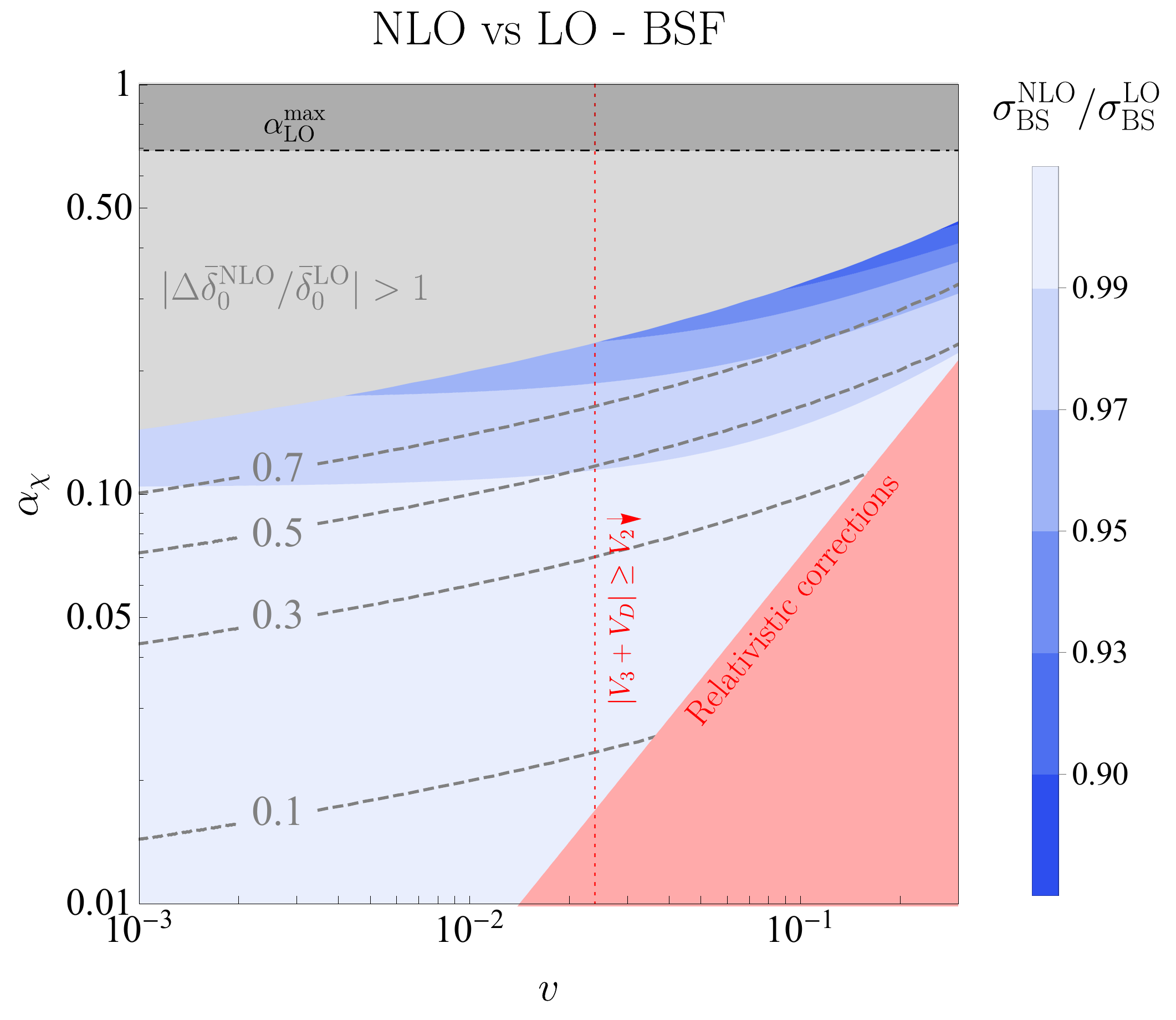}
         \includegraphics[width=0.51\textwidth]{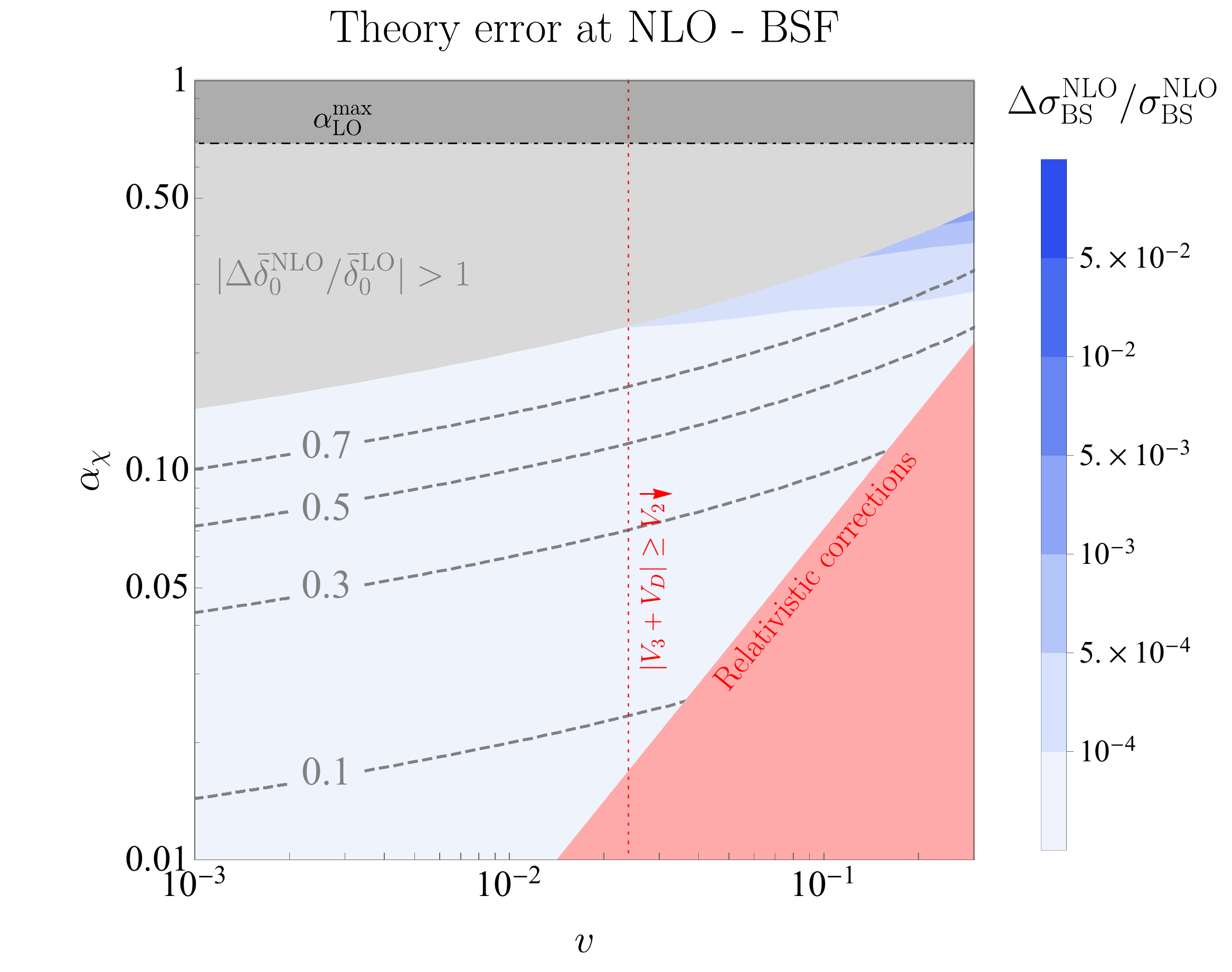}
\caption{Behavior of the UV NLO corrections to BSF described in Sec.~\ref{sec:UV} with $x_{\text{cut}}=1$ and Appendix \ref{app:VPM}. The red lines exhibit the power counting on the relativistic corrections discussed below Eq.~\eqref{eq:NLO_UV}. On the right of the {\bf red} dashed line, the NLO potential cannot be approximated with a $1/r^2$ correction to the Coulomb potential while on the {\bf red} shaded region the corrections to the kinetic energy become important and the non-relativistic description breaks down. The {\bf gray dashed} lines show the NLO contribution to the UV phase in the Born approximation as defined in Eq.~\eqref{eq:NLOphaseshift}, normalized with respect to the LO one. In the {\bf gray} shaded area the NLO contribution becomes larger than the LO one and the calculability of the phase breaks down. {\bf Left:} The {\bf blue} contours show the behavior of the mean value of the NLO $\sigma_{\rm BS}$ normalized with respect to the LO $\sigma_{\rm BS}$. {\bf Right:} The {\bf blue} contours show the behavior of the theory error on the NLO $\sigma_{\rm BS}$ due to the uncertainty on the UV phase determination as defined in Eq.~\eqref{eq:SEerrorphase}.}
\label{fig:UV_plots_bsf}
\end{figure}

In this Section, we briefly comment on the sensitivity of BSF to the UV dynamics. As we already mentioned in the main text, BSF is expected to be only weakly affected by the UV details of the theory. This is due to the fact that, while the relevant scale for BSF is the Bohr radius $a_B=2/\alpha_\chi M_\chi$, the UV corrections to the LO Coulombian dynamics are relevant only for $r<1/M_\chi$. Besides, because of the electric dipole selection rules, the angular momentum of at least one between the BS and the scattering state wave function must be non-zero, so that the centrifugal potential in the Schrodinger equation further screens the UV potentials. In Fig. \ref{fig:UV_plots_bsf} we show the equivalent of Fig. \ref{fig:UV_plots} for BSF. As we can see, even close to the boundary where the UV phase becomes incalculable, the UV corrections to $\sigma_{\rm BS}$ is less than 10 \% of its value at LO. Moreover, the uncertainty in the UV phase induces sub-percent corrections to $\sigma_{\rm BS}$ at NLO.

\subsection{Variable Phase Method for Bound State Formation}\label{app:VPM}

We review the Variable Phase Method (VPM) introduced in \cite{Ershov:2011zz} (see also \cite{Beneke:2014gja,Asadi:2016ybp,Mahbubani:2020knq}) to determine the initial scattering wave function to be plugged in the overlap integrals in \eqref{eq:overlap}. We describe a straightforward generalization to discontinuous potentials, which allows us to compute the BSF cross-section in the presence of the regularized potential in Eq.~\eqref{eq:reg_pot}.

\paragraph{Continuous potential.} First we review the VPM in the case of a continuous potential, assuming for simplicity a single scattering channel. This case is the case, for example, of both the LO and the NLO potentials with IR contributions in the scalar DM scenario. The generalization to multiple channels, like in the case of DM belonging to a multiplet or when spin-dependent potentials are present, like for fermionic DM, is straightforward.  Consider the scattering problem described by the following \schr equation

\begin{equation}\label{eqapp:schr}
    u''(z)+\left(1+\frac{l(l+1)}{z^2}\right)u(z)=\frac{4\mathcal{V}(z)}{ \vrel^2} u(z)\, ,
\end{equation}
where $u(z)=zR_p(z)$ is the reduced wave function and $z=M_\chi \vrel r/2$ and $\mathcal{V}(z)=V(r)/M_\chi$. The VPM consists in writing the solution to the previous equation as a linear combination of the solutions to the free \schr equation
\begin{equation}
    \begin{pmatrix}
 f_l(z)\\
 g_l(z)
    \end{pmatrix}''+\left(1+\frac{l(l+1)}{z^2}\right)\begin{pmatrix}
 f_l(z)\\
 g_l(z)
    \end{pmatrix}=0\, ,
\end{equation}
where $f_l$ and $g_l$ are the regular and irregular solutions, repsectively, and are normalized as
\begin{equation}\label{eqapp:norm}
    f_l(z)g_l'(z)-g_l(z)f_l'(z)=-1\ ,
\end{equation}
so that
\begin{equation}\label{eqapp:fg}
    f_l(z)=zj_l(z),\quad g_l(z)=-z(y_L(z)-ij_l(z))\, ,
\end{equation}
where $j_l$ ($y_l$) is the (ir)regular $l^{\text{th}}$ spherical Bessel function. We can thus write the solution to Eq.~\eqref{eqapp:schr} as
\begin{equation}
    u(z)=f_l(z)\alpha(z)-g_l(z)\beta(z)\, ,
\end{equation}
where $\alpha(z)$ and $\beta(z)$ are unknown functions. Having doubled the number of unknowns, we impose the  constraint
\begin{equation}\label{eqapp:constraint}
    f_l(z)\alpha'(z)=g_l(z)\beta'(z)\ .
\end{equation}
Finally, in order to improve the numerical convergence, we introduce the function
\begin{equation}
    N(z)=\left(f_l(z)-g_l(z)O(z)\right)g_l(z),\quad O(z)\equiv \frac{\beta(z)}{\alpha(z)}\ ,
\end{equation}
so that $u=N\alpha/g_l$. In this way, the original second-order \schr equation Eq.~\eqref{eqapp:schr} is split into two first order equations

\begin{equation}\label{eqapp:VPM}
\begin{aligned}
    &N'(z)=1+2\frac{g'_L(z)}{g_l(z)}N(z)-\frac{4\mathcal{V}(z)}{ \vrel^2}N(z)^2,\\
    &\alpha'(z)=\frac{4\mathcal{V}(z)}{\vrel^2}N(z)\alpha(z).
    \end{aligned}
\end{equation}
Requiring that close to the origin $u(z\rightarrow 0)=z^{l+1}/(2l+1)$ is equivalent to the following boundary conditions for $N(z)$ and $\alpha(z)$
\begin{equation}
    N(z\rightarrow 0)=\frac{z}{2l+1},\quad \alpha(0)=(2l-1)!!\ .
\end{equation}
An equivalent, numerically more stable boundary condition for $\alpha(z)$ is $\alpha(z\rightarrow\infty)=1$.\\

\paragraph{Discontinuous potential} Here we generalize the VPM method to the case of the regularized potential in Eq.~\eqref{eq:reg_pot}

\begin{equation}
\mathcal{V}_{\text{reg}}(x)=\left\{
\begin{aligned}
&-\vcut(
v,x_{\text{cut}}),\quad x<\xcut\\
&\mathcal{V}_{\text{NLO}}^{\text{UV}}(x),\qquad\quad\ \ x>\xcut
\end{aligned}\right.\ ,
\end{equation}
The \schr equation inside the potential well reads:
\begin{equation}
    Z_\chi u''_<(z)+\left(1+Z_\chi\frac{l(l+1)}{z^2}\right)u_<(z)=-\frac{4\vcut}{ \vrel^2} u_<(z)\, .
\end{equation}
In order to make the matching of $u_<(z)$ to the solution $u>(z)$ of the \schr equation outside the well, we write:
\begin{equation}
    u_<(z)=f_l^<(z)\alpha_<(z)-g_l^<(z)\beta_<(z)\, ,
\end{equation}
where $f_l^<(z)$ and $g_l^<(z)$ solve
\begin{equation}
    Z_\chi\begin{pmatrix}
 f_l^<(z)\\
 g_l^<(z)
    \end{pmatrix}''+\left(1+Z_\chi\frac{l(l+1)}{z^2}\right)\begin{pmatrix}
 f_l^<(z)\\
 g_l^<(z)
    \end{pmatrix}=0\ .
\end{equation}
We keep the same normalization and constraint as in Eq.~\eqref{eqapp:norm} and Eq.~\eqref{eqapp:constraint}, so that
\begin{equation}
    f_l^<(z)=Z_\chi^{\frac{1}{4}}zj_l\left(\frac{z}{\sqrt{Z_\chi}}\right),\quad g_l^<(z)=-Z_\chi^{\frac{1}{4}}z\left(y_l\left(\frac{z}{\sqrt{Z_\chi}}\right)-ij_l\left(\frac{z}{\sqrt{Z_\chi}}\right)\right)\ .
\end{equation}
Equivalently to the continuous case, we define the function $N_<$ inside the well as
\begin{equation}
    N_<(z)=\left(f^<_L(z)-g^<_L(z)O_<(z)\right)g^<_L(z),\quad O_<(z)\equiv \frac{\beta_<(z)}{\alpha_<(z)}\, ,
\end{equation}
where now $N_<$ and $\alpha_<$ solve
\begin{equation}
\begin{aligned}
    &N'_<(z)=1+2\frac{g'^<_L(z)}{g^<_L(z)}N_<(z)+\frac{4\vcut}{ Z_\chi\vrel^2}N^2_<(z)\ ,\\
    &\alpha'_<(z)=-\frac{4\vcut}{Z_\chi\vrel^2}N_<(z)\alpha_<(z)\ .
    \end{aligned}
\end{equation}
The matching of $N_<$ and $\alpha_<$ to the corresponding functions defined outside the well, $N_>$ and $\alpha_>$ is obtained by requiring the continuity of $u(z)$ and $u'(z)$ across the boundary of the well itself. This leads to
\begin{equation}
    O_>=\left.\frac{f_lf'^<_L-f_l'f_l^<+O_<(g_l^<f_l'-g'^<_Lf_l)}{g_lf'^<_L-g_l'g_l^<+O_<(g_l^<g_l'-g'^<_L<g_l)}\right|_{z=\vrel\xcut/2}\ ,
\end{equation}
where $f_l$ and $g_l$ are defined in Eq.~\eqref{eqapp:fg} while $O_<(\vrel\xcut/2)$ is determined solving the equation for $N_<$ with initial condition $N_<(z\rightarrow 0)=z^{l+1}/(2l+1)$. Finally, the matching of $\alpha_<$ is simply
\begin{equation}
    \left.\frac{N_<\alpha_<}{g_l^<}\right|_{z=\vrel\xcut/2}= \left.\frac{N_>\alpha_>}{g_l}\right|_{z=\vrel\xcut/2}\ ,
\end{equation}
where $\alpha_>$ solves Eq.~\eqref{eqapp:VPM} with the same boundary condition $\alpha_>(z\rightarrow\infty)=1$.

\end{document}